\documentclass[twocolumn,           
               showpacs,            
               preprintnumbers,     
               aps,                 
               prd,                 
               letter,             
               groupedaddress,  
               nofootinbib,         
               tightenlines,        
               floats,floatfix      
               ]{revtex4}

\usepackage{graphicx}
\usepackage{subfigure}
\usepackage{latexsym}
\usepackage{amsmath,amssymb}        
\usepackage[draft=false]{hyperref}
\usepackage{threeparttable}

\renewcommand{\(}{\left(}
\renewcommand{\)}{\right)}

\newcommand{\reffig}[1]{Fig.~\ref{#1}}

\input colordvi

\begin{document}



\title{Quintessence reconstructed: new constraints and tracker viability}
\author{Martin Sahl\'en}
\affiliation{Astronomy Centre, University of Sussex, Brighton BN1 9QH,
  United Kingdom}
\author{Andrew R.~Liddle}
\affiliation{Astronomy Centre, University of Sussex, Brighton BN1 9QH,
  United Kingdom}
\author{David Parkinson}
\affiliation{Astronomy Centre, University of Sussex, Brighton BN1 9QH,
  United Kingdom}
\date{\today}
\pacs{98.80.-k \hfill astro-ph/0610812}
\preprint{astro-ph/0610812}


\begin{abstract}
We update and extend our previous work reconstructing the potential
of a quintessence field from current observational data. We extend
the cosmological dataset to include new supernova data, plus
information from the cosmic microwave background and from baryon
acoustic oscillations. We extend the modelling by considering Pad\'e
approximant expansions as well as Taylor series, and by using
observations to assess the viability of the tracker hypothesis. We
find that parameter constraints have improved by a factor of two,
with a strengthening of the preference of the cosmological constant
over evolving quintessence models. Present data show some signs,
though inconclusive, of favouring tracker models over non-tracker
models under our assumptions.
\end{abstract}

\maketitle


\section{Introduction}

The nature of dark energy in our Universe remains unknown, and is
likely to be the subject of intense observational attention over the
coming decade \cite{DETF}. While a pure cosmological constant remains
the simplest interpretation of present data, a leading alternative
possibility is the quintessence paradigm, whereby the observed
acceleration is driven by the potential energy of a single
canonically-normalized scalar field \cite{quint} (for extensive
reviews of dark energy see Ref.~\cite{reviews}). In this paper, we
work under the assumption that quintessence is a valid description of
observational data (an assumption to be tested separately), and seek
to impose optimal constraints on the model via exact numerical
computation. Our work provides an implementation of quintessence
potential reconstruction, a subject developed by several authors
\cite{rec, dalyrec,oldppr}, and by assuming a particular physical
model for dark energy is distinct from parameterized equation of state
methods for reconstructing dark energy.

In a previous paper \cite{oldppr}, we carried out a direct
reconstruction of the quintessence potential based on the supernova
type Ia (SNIa) luminosity--redshift measurements made/collated by
Riess {\it et al.}~\cite{riess}. The present paper updates and extends
that work in three ways:
\begin{enumerate}
\item We include additional data coming from cosmic microwave
  background (CMB) anisotropies \cite{wmap3} and baryon acoustic
  oscillations \cite{sdssbao}, as well as using newer supernova data
  from the SuperNova Legacy Survey (SNLS) \cite{snls}. We do not use
  constraints from the growth rate of structure, which are not yet
  competitive with the data we do use.
\item Where previously we approximated the quintessence potential via
  a Taylor series, we now additionally explore use of Pad\'e
  approximant expansions in order to test robustness under choice of
  expansion. \vspace*{20pt}
\item By studying the dynamical properties of models permitted by the
  data, we assess whether current observations favour or disfavour the
  hypothesis that the quintessence field is of tracker form, hence
  potentially addressing the coincidence problem.
\end{enumerate}

As we were completing this paper, a closely-related paper was
submitted by Huterer and Peiris \cite{HP}, who also reconstruct
quintessence potentials from a similar compilation of current data.
Although phrased in the language of flow equations, their approach,
like ours here and in Ref.~\cite{oldppr}, amounts to fitting the
coefficients of a Taylor expansion of the potential. They do not
consider Pad\'e approximants. Their approach implies different
priors for the parameters than the ones used in this paper, and they
treat the scalar field velocity a little differently. Our results
appear in good agreement, in particular our determination that
present data mildly favour tracker models over non-tracker models
concurring with their conclusion that freezing models are mildly
preferred to thawing ones (in the terminology of Ref.~\cite{CL}).

\section{Formalism}

\subsection{Cosmological model}

We quickly review the set-up of Ref.~\cite{oldppr}, which is
conceptually straightforward. We assume that the quintessence field
$\phi$ has a potential $V(\phi)$, which we expand in a series about
the present value of the field that is taken (without loss of
generality) to be zero. The quintessence field obeys the equation
\begin{equation}
\ddot{\phi} + 3 H \dot{\phi} = - \frac{dV}{d\phi} \,,
\end{equation}
with the Hubble parameter $H$ given by the Friedmann equation
\begin{equation}
H^2 = \frac{8\pi G}{3} \left(\rho_{{\rm m}} + \rho_{\rm \phi}
\right) \,.
\end{equation}
Here $\rho_{{\rm m}}$ is the matter density and $\rho_{\rm \phi} =
\dot{\phi}^2/2 + V(\phi)$ the quintessence density. We assume
spatial flatness throughout (as motivated by CMB measurements and
the inflationary paradigm), though the generalization to the
non-flat case would be straightforward. Since then $\Omega_{\rm m} +
\Omega_{\phi} = 1$ we have the present boundary condition
\begin{equation}
\label{eq:phidot} \dot{\phi}_{\rm 0} = \pm
\sqrt{2\left[(1-\Omega_{\rm m})\rho_{\rm c,0} - V(\phi_0) \right]} \,,
\end{equation}
where subscript `0' indicates present value, and $\rho_{{\rm c}}$ is
the critical density.  An important quantity, which determines the
cosmological effects we consider from the quintessence field, is the
equation of state
\begin{equation}
w_{\phi} \equiv \frac{p_{\phi}}{\rho_{\phi}} =
\frac{\dot{\phi}^2/2 - V(\phi)}{\dot{\phi}^2/2 + V(\phi)} \,.
\end{equation}
The priors we assume for our cosmology are
\begin{eqnarray}
    \Omega_{\rm total} & = & 1 \,, \\
    \Omega_{\rm m} & \ge & 0 \,, \\
    \Omega_{\rm kin} & \le & 1 \,, \\
    \Omega_{\rm kin}(z\ge1) & < & 0.5 \,.
\end{eqnarray}
where $\Omega_{\rm kin} = 8\pi G\dot{\phi}^2/6H^2$ is the fraction of
critical energy density in field kinetic energy density.  The last
condition is a means of encoding that the field should not interfere
too much with structure formation (as we do not use data sensitive to
that), and is discussed further in our previous paper
\cite{oldppr}. The constraint on $\Omega_{\rm kin}$ is in practice
applied up to the highest redshift for which we have data points,
i.e.~using CMB information $z=1089$. When we use supernova data only,
the upper limit is $z=2$, as in our previous study.

\subsection{Parameterizations and priors}

To explore the space of potentials, we need to assume some
functional form for the potential. We choose two classes of
expansions, a Taylor series, and a Pad\'e series, to parameterize
the potential function $V(\phi)$. In the absence of a theoretical
bias for the functional form of the potential, these expansions seem
suitably general and simple to provide a reasonably fair sampling of
the space of potential functions.

\subsubsection{Taylor series}

As in our previous study, we use a Taylor series to model the
potential $V(\phi)$ as
\begin{equation}
V(\phi) = V_0 + V_1\phi + V_2\phi^2 + \ldots
\end{equation}
where $\phi$ is in units of the reduced Planck mass $M_{\rm P}$ with
$\phi$ presently zero. We will refer to a constant potential with
non-zero kinetic energy allowed as a \emph{skater} model, after
Linder in Ref.~\cite{linderskater}.

We put the following flat priors on the parameters:
\begin{equation}
    V_0  \ge 0 \,, \quad
    |V_1|  \le  2 \,, \quad
    |V_2| \le  5 \,.
\end{equation}
These priors are irrelevant for parameter estimation, as they are
significantly broader than the high-likelihood region (this also
applies to the corresponding priors for Pad\'e series below).
However, to assess how favoured tracker behaviour is, we do need to
put some limits, so that we can sample a finite region of the prior
parameter space (see further in Section~\ref{sect:trvsnontr}).

\subsubsection{Pad\'e series}

In addition to the Taylor series expansion, in this paper we also
use Pad\'e approximant expansions in order to test the robustness of
results to the method used. Pad\'e approximants are rational
functions of the form
\begin{equation}
R_{M/N}(\phi) = \frac{\sum_{i=0}^M a_i \phi^i}{1+\sum_{j=1}^N b_j
\phi^j} \,,
\end{equation}
that can be used to approximate functions. These approximants
typically have better-behaved asymptotics, i.e.~stay closer to the
approximated function, than Taylor expansions because of their
rational structure. An extensive expos\'e on Pad\'e approximants can
be found in Ref.~\cite{pade}. For our study, we will assume
\begin{equation}
V(\phi) = R_{M/N}(\phi) \,,
\end{equation}
where again $\phi$ is in units of $M_{\rm P}$ with $\phi$ presently
zero. Specifically, we use Pad\'e series $R_{0/1}$, $R_{1/1}$ and
$R_{0/2}$, as these form an exhaustive set of lowest order and
next-to-lowest order non-trivial expansions with two or three
parameters. Higher orders are unmotivated given the known difficulty
for data to constrain more than two dark energy/quintessence evolution
parameters \cite{linderhowmany,oldppr,Maor:2002rd} (as will also be
evident from our results).

Pad\'e series have poles, but, as will be discussed in the Results
Section, data constrains models so that the presence of poles is not
felt.

To enable comparison between our results for the two different
parameterization classes, the priors for the Pad\'e series case are
set by evaluating the MacLaurin expansion of the Pad\'e series,
identifying the order coefficients, and using the Taylor-series
priors for those, i.e.
\begin{eqnarray}
    a_0 & = & V_0 \,,\\
    a_1 - a_0b_1 & = & V_1 \,,\\
    b_1(a_0b_1 - a_1) - a_0b_2 & = & V_2 \,.
\end{eqnarray}
This does not limit us to a finite region, so we additionally
require $|b_1| \le 2$.

\subsection{Tracker potentials}

Cosmological tracker potentials/solutions have been studied in detail
by numerous authors
\cite{quint,reftrack,steintrack,bludman,linderpaths}. These potentials
are such that the late-time evolution of the field can be essentially
independent of initial conditions, thus providing a possible solution
to the coincidence problem. This behaviour is achieved through a type
of dynamical attractor solution, and the conditions for it to be
possible given a particular potential have been given and studied in
detail by Steinhardt {\it et al.}  \cite{steintrack}. Defining $\Gamma
\equiv V''V/V'^2$, where prime denotes a derivative with respect to
the field, the two sufficient conditions for a potential to possess a
tracker solution are
\begin{eqnarray}
 \label{eq:gammareq}
 \Gamma &> &1 - \frac{1-w_{\rm b}}{6+2w_{\rm b}} \,, \\
 \left|\Gamma^{-1}\frac{d \Gamma}{d\ln a}\right| & = &
\left|\frac{d\phi}{d\ln a}\(\frac{V'}{V} + \frac{V'''}{V''} -
2\frac{V''}{V'}\)\right| \ll 1 \,. \quad
\end{eqnarray}
The first of these conditions ensures convergence to the tracker
solution (i.e.~perturbations away from it are suppressed), and the
second ensures an adiabatic evolution of the field that is necessary
for the first condition to be applicable (and is what one would
expect of a function that is to maintain a dynamical attractor
independent of initial conditions).

\begin{figure}[t]
\includegraphics[width=0.8\linewidth]{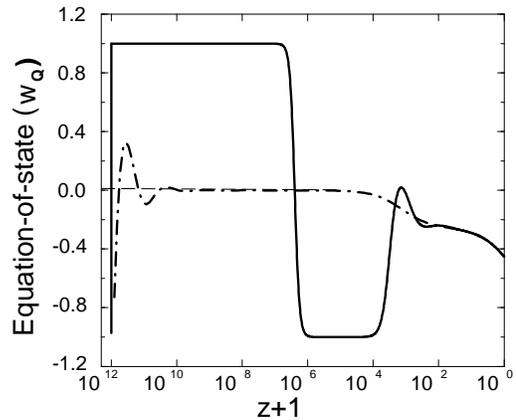}
\caption{Examples of the behaviour of the equation of state (here
called $w_{\rm Q}$) for a tracker potential. The oscillating curves
correspond to higher (solid) and slightly lower (dash-dotted) initial
conditions at high redshift for $\rho_{\phi}$ compared to the tracker
solution value. The initial velocity at high redshift is assumed to be
zero. The amplitude of oscillations in $w(z)$ around the tracker
solution (thin-dashed curve originating at $w_{\rm Q}=0$) decays
exponentially with decreasing $\ln(1+z)$, and the evolution thus
approaches the tracker solution regardless of the different initial
conditions.  Although not directly corresponding to our models, the
figure illustrates qualitatively the tracker property. Reproduced from
Ref.~\cite{steintrack}.}
\label{fig:trackex}
\end{figure}

If these conditions are fulfilled, the field will eventually approach
the tracker solution (unless the initial quintessence energy density
is too low), and the equation of state will then evolve according to
\begin{equation}
  \label{eq:wtracker}
  w_{\phi} \approx w_{\rm tracker} = \frac{w_{\rm b} -
  2(\Gamma-1)}{1+2(\Gamma-1)} \,,
\end{equation}
possibly breaking away from the tracker solution if either of the
conditions later become violated. In assessing whether tracking is
taking place, one also has to check whether the actual evolution on
the tracker potential corresponds closely to the tracker solution.
An illustration of tracker behaviour can be seen in
Fig.~\ref{fig:trackex}.

We additionally impose the condition $w_{\phi} < w_{\rm b}$, where
$w_{\rm b}$ is the background energy density. This is to ensure a
possible solution of the coincidence problem by having the dark
energy density grow with respect to the matter. This third condition
is usually avoided by specifying the tracker condition as $\Gamma>1$
rather than Eq.~(\ref{eq:gammareq}). The reason for not choosing
$\Gamma>1$ as our condition is related to our numerical treatment,
and is discussed further in Section~\ref{sect:identtrack}.

As we need a non-zero second derivative of the potential with
respect to the field for $\Gamma$ to fulfil the tracker conditions,
we restrict ourselves to the quadratic potential and the Pad\'e
series for the tracker viability analysis.

\section{Observables}

The observables used are essentially geometric and are hence related
to the comoving distance for an FRW cosmology described by the
parameter vector $\mathbf{\Theta}$, given by
\begin{equation}
r(z; \mathbf{\Theta}) = H_0^{-1}\int_0^z \frac{{\rm d}z'}{E(z';
\mathbf{\Theta})} \,
\end{equation}
where
\begin{equation}
E(z; \mathbf{\Theta}) = \left[ \Omega_{\rm m}(1+z)^3 + (
1-\Omega_{\rm m}) e^{F(z; \mathbf{\Theta})} \right]^{1/2}
\end{equation}
and
\begin{equation}
F(z; \mathbf{\Theta}) =  3 \int_0^{z} \left(1+w_{\phi}(z';
\mathbf{\Theta})\right) {\rm d}\ln(1+z') \,.
\end{equation}
In accordance with our assumptions, these expressions assume zero
curvature and that quintessence and non-relativistic matter are the
only relevant components for the redshifts we consider.

We have not included growth-of-structure observations, which are not
yet competitive with the measures we do use (see
e.g.~Ref.~\cite{wangkratochvil} for a directly-comparable example).

\subsection{SNIa luminosity--redshift relation}

The luminosity distance is given by
\begin{equation}
\label{eq:lumdistz} d_{\rm L}(z; \mathbf{\Theta}) =
\frac{\mathcal{D}_{\rm L}(z; \mathbf{\Theta})}{H_0} = (1+z)r(z;
\mathbf{\Theta}) \,.
\end{equation}
The apparent magnitude $m(z; \mathbf{\Theta})$ of a type Ia
supernova can be expressed as
\begin{equation}
m(z; \mathbf{\Theta}) = M + 5\log_{10} \frac{d_{\rm L}(z;
\mathbf{\Theta})}{\rm Mpc} + 25 \,,
\end{equation}
where $M$ is the absolute magnitude of SNIa (supposing they are
standard candles). This can be rewritten as
\begin{equation}
m(z; \mathbf{\Theta}) = \mathcal{M} + 5\log_{10} \mathcal{D}_{\rm
L}(z; \mathbf{\Theta}) \,,
\end{equation}
where $\mathcal{M} = M - 5\log_{10} \left(H_{\rm 0}\,\,{\rm Mpc}
\right) + 25 = M - 5\log_{10}(h_{70}) + 43.16$ [where $h_{70} =
H_0/(70\, {\rm km/s/Mpc})$]. Note that some authors define this quantity
somewhat differently.

We use the 115 measurements of $m(z)$ measured/compiled by the SNLS
team \cite{snls}, covering the redshift range $z=0.015$ to $z=1.01$. The
observed magnitudes (indexed by $i$) are given by
\begin{equation}
m_i = m^*_{{\rm B}, i} + \alpha(s_i-1) - \beta c_i
\end{equation}
where $m^*_{\rm B}$ is the rest-frame B-band magnitude at maximum
B-band luminosity, and $s$ and $c$ are the shape and color
parameters. These are derived from the light-curve fits and are
reported by the SNLS team. The parameters $\alpha$ and $\beta$ are
free parameters and should be varied in cosmological fits. However, as
they are independent of cosmology \cite{fouchezpc}, we fix them to the
SNLS best-fit values
\begin{eqnarray}
\alpha & = & 1.52 \pm 0.14 \,, \\
\beta & = & 1.57 \pm 0.15 \,,
\end{eqnarray}
without introducing any bias, and include their uncertainty in the
magnitude uncertainties we use.

For comparison
to our previous paper where the parameter $\eta$ is used, the
parameter $\mathcal{M} = M^{*}_{\rm Riess} - \eta$, with $M^{*}_{\rm
Riess}$ the estimate of intrinsic supernova magnitude in Riess {\it et al.}
\cite{riess}.

\subsection{CMB peak-shift parameter}

The CMB peak-shift parameter \cite{peakshift}
\begin{equation}
\mathcal{R}(z_{\rm dec}; \mathbf{\Theta}) = \sqrt{\Omega_{\rm m}}
H_0 r(z_{\rm dec}; \mathbf{\Theta})
\end{equation}
measures an overall linear shift of the CMB power spectrum in
multipole space, induced by the effect $\Omega_{\phi}$ has on the
angular-diameter distance to the surface of last scattering at
$z=z_{\rm dec}$. The position of the first power spectrum peak is
essentially a measure of this distance.

We use the recent WMAP3 data \cite{wmap3} as analyzed by Wang and
Mukherjee \cite{wangr}, who found
\begin{equation}
\mathcal{R}(z_{\rm dec}=1089) = 1.70 \pm 0.03 \,.
\end{equation}

\subsection{Baryon acoustic peak}

The standard Big Bang scenario predicts that close to the surface of
last scattering, baryons and photons act as a fluid with acoustic
oscillations from the competition between gravitational attraction and
radiation pressure. As the photons decouple, these acoustic
oscillations should be frozen in the baryon and dark matter
distributions. One would thus expect an excess of power in the power
spectrum of luminous matter at a scale corresponding to the acoustic
scale at last scattering (see e.g.~Ref.~\cite{baoth} and references
therein). Independent first detections of this baryon acoustic peak
were made by the Sloan Digital Sky Survey (SDSS) \cite{sdssbao} and
the 2dF galaxy redshift survey \cite{2dfbao}. The SDSS team defined a
distance quantity
\begin{equation}
A(z_{\rm BAO}; \mathbf{\Theta}) = \sqrt{\Omega_{\rm m}} \left(
\frac{H_0^2 r^2(z_{\rm BAO}; \mathbf{\Theta})}{z^2_{\rm BAO}E(z_{\rm
BAO}; \mathbf{\Theta})} \right)^{1/3} \,,
\end{equation}
which we will use for our analysis. The measurement (independent of
dark energy model) from the SDSS luminous red galaxy power spectrum
is \cite{sdssbao}
\begin{equation}
A(z_{\rm BAO}=0.35) = 0.469 \left(\frac{n_{\rm
S}}{0.98}\right)^{-0.35} \pm 0.017 \,,
\end{equation}
which, assuming the WMAP3 mean value $n_{\rm S}=0.95$ \cite{wmap3},
yields $A(z=0.35) = 0.474 \pm 0.017$.

\section{Data analysis}

\subsection{Parameter estimation}

The parameter space we study will be
\begin{equation}
\mathbf{\Theta} = (\mathcal{M}, \dot{\phi}_{\rm 0}, \mbox{potential
parameters}) \,,
\end{equation}
and we will consistently let $D$ denote the number of free
parameters in a model.  The parameter estimation is carried out
using an MCMC approach, as outlined in our previous paper
\cite{oldppr}. The posterior probability of the parameters
$\mathbf{\Theta}$, given the data and a prior probability
distribution $\Pi(\mathbf{\Theta})$, is
\begin{equation}
P(\mathbf{\Theta} | {\rm data}) = \frac{1}{{\cal Z}}
e^{-\left(\chi_{\rm SNIa}^2(\mathbf{\Theta})+\chi_{\rm
CMB}^2(\mathbf{\Theta})+\chi_{\rm BAO}^2(\mathbf{\Theta})\right)/2}
\Pi(\mathbf{\Theta}) \,,
\end{equation}
where
\begin{eqnarray}
\chi_{\rm SNIa}^2(\mathbf{\Theta}) & = & \sum_{i=1}^{N_{\rm SNIa}}
\frac{\left(m_i
- m(z_i; \mathbf{\Theta})\right)^2}{\sigma_i^2} \,, \\
 \chi_{\rm CMB}^2(\mathbf{\Theta}) & = & \frac{\left(\mathcal{R}_{\rm
     obs} - \mathcal{R}(z_{\rm dec};
\mathbf{\Theta})\right)^2}{\sigma_{\mathcal{R}}^2} \,, \\
\chi_{\rm BAO}^2(\mathbf{\Theta}) & = & \frac{\left(A_{\rm obs} -
A(z_{\rm BAO}; \mathbf{\Theta})\right)^2}{\sigma_{\rm A}^2} \,.
\end{eqnarray}
Here, we sum over all $N_{\rm SNIa}$ data points for the SNIa data,
and ${\cal Z}~=~\int {\cal L}({\rm data}|\mathbf{\Theta})
\Pi(\mathbf{\Theta}) {\rm d}\mathbf{\Theta}$ is a normalization
constant, irrelevant for parameter fitting.  Overall, we have
115(SNIa)+1(CMB)+1(BAO) data points.

\subsection{Model selection}

Separate from the question of parameter estimation is the question
of parameter necessity, i.e. model selection. We again employ an
approximate model selection criterion, the Bayesian Information
Criterion (BIC) \cite{Schwarz,Lid}, given by
\begin{equation}
\label{eq:bic}
 \mathrm{BIC} = -2\ln{\cal L}_{{\rm max}} + D \ln N
\,,
\end{equation}
where ${\cal L}_{{\rm max}}$ is the likelihood of the best-fitting
parameters for that model, $D$ the number of model parameters, and $N$
the number of datapoints used in the fit. Models are ranked with the
lowest value of the BIC indicating the preferred model. A difference
of two for the BIC is regarded as positive evidence, and of six or
more as strong evidence, against the model with the larger value
\cite{jeff,Muk98}.
The BIC has also been deployed for dark energy model selection in
Ref.~\cite{darkBIC}.

\subsection{Tracker viability}

\subsubsection{Identifying tracker solutions}
\label{sect:identtrack}

To classify general scalar field evolutions as coming from a tracker
potential capable of solving the coincidence problem or not, we need
to test for both tracker conditions and whether the field evolves
according to the tracker solution. As these conditions are
approximate in nature, we must specify some $\epsilon\ge 0,
\delta\ge 0$ such that if
\begin{eqnarray}
  \Gamma & >& 1 - \frac{1-w_{\rm b}}{6+2w_{\rm b}} \,, \\
  \left|\Gamma^{-1}\frac{d \Gamma}{d\ln a}\right| &<& \epsilon \,, \\
  \left|w_{\phi}-w_{\rm tracker}\right|& <&
  \delta \,, \\
   w_{\phi}&<&w_{\rm b} \,,
\end{eqnarray}
are all fulfilled for some range in redshift over which we require the
field to be in the tracker solution, the potential is classified as a
tracker potential. To provide a satisfactory solution to the
coincidence problem, the field should have $w_{\phi} < w_{\rm b}$
while in the tracker solution.  This condition is automatically
satisfied if the tracker conditions are fulfilled with $\Gamma>1$ and
the field is in the tracker solution. However, in our analysis there
is some room for fields with $w_{\phi} \ge w_{\rm b}$, since the field
is allowed to deviate slightly from the tracker solution, and we also
consider $\Gamma>1-(1-w_{\rm b})/(6+2w_{\rm b})$ as tracking rather
than $\Gamma>1$ that is typically used. Cases satisfying the former
but not the latter are generally disfavoured because they would
correspond to $w_{\phi}>w_{\rm b}$ in the tracker solution and hence
not be very successful for solving the coincidence problem. In our
set-up this is not necessarily true, and this is the reason for not
choosing the more commonly-used latter criterion. Instead, we ensure a
solution to the coincidence problem by enforcing $w_{\rm \phi}<w_{\rm
b}$. In particular, we require $\Gamma > 5/6$ and $w_{\phi} < 0$ since
we are concerned with the matter-dominated epoch.

Note that we are not connecting our analysis directly with any
specific particle physics model and its initial conditions at early
times, and assessing whether the present-time observables are highly
insensitive to variations in those initial conditions. We are only
addressing the question whether the (essentially late-time)
evolution of quintessence is more consistent with such a class of
tracker potentials, or with a class that does not have such
behaviour. As the shape of the potential at high redshifts is
almost unconstrained by data (see also e.g.~Ref.~\cite{dalyrec}), we
adopt the viewpoint that a suitable \emph{true} tracker potential
with insensitivity to initial conditions can always be made to
coincide with our low-redshift behaviour.

\subsubsection{Tracker or Non-Tracker?}
\label{sect:trvsnontr}

To assess whether models which exhibit tracker solution behaviour are
favoured by data over models which do not, we need some quantity to
measure this preference. A well-defined and well-motivated quantity is
provided within the framework of Bayesian model selection
\cite{Lid,jeff,bayesms}, where the Bayes factor
\begin{equation}
B_{12} \equiv \frac{P({\rm D}|M_1)}{P({\rm D}|M_2)} =
\frac{P(M_1|{\rm D})}{P(M_2|{\rm D})} \frac{\Pi(M_2)}{\Pi(M_1)} \,,
\end{equation}
simply the relative power of Model 1 ($M_1$) over Model 2 ($M_2$) in
explaining the observed data $D$ given the prior model probabilities
$\Pi(M_1)$ and $\Pi(M_2)$, can be used to perform this type of
comparison.

For the purposes of assessing the viability of tracker solutions for
explaining the observed data, we will define the following models:
\begin{eqnarray}
M_1 & = & \{\mbox{$V$ is a tracker potential}\} \,, \\
M_2 & = & \{\mbox{$V$ is not a tracker potential}\} \,.
\end{eqnarray}
As these two models are disjoint subsets of the model space, the
Bayes factor can be estimated from Monte Carlo Markov chains:
letting $f_{\rm post}$ be the fraction of chain elements from the
posterior distribution satisfying the tracker criteria, and $f_{\rm
pri}$ the corresponding fraction for the prior distribution, the
Bayes factor is given by
\begin{equation}
\label{eq:bayes}
 B_{12} \approx \frac{f_{\rm post}(1-f_{\rm
pri})}{f_{\rm pri}(1-f_{\rm
    post})} \,,
\end{equation}
since the fractions of tracker and non-tracker chain elements must
sum to one for both prior and posterior. In the limit of equal
fractions in prior and posterior, $B_{12} = 1$, whereas in the limit
of complete suppression of tracker models in the posterior (so that
$f_{\rm post}=0$) we have $B_{12} = 0$ in which case Model 2 is
infinitely favoured over Model 1.

\begin{table}
\begin{tabular}{|c|c|}
    \hline
    $\ln(B_{12})$ & Evidence against Model 2 \\
    \hline\hline
    $0 - 1$ & Worth only a bare mention \\
    $1 - 2.5$ & Positive evidence \\
    $2.5 - 5$ & Strong evidence \\
    $>5$ & Decisive evidence \\
    \hline
\end{tabular}
\caption{The Jeffreys evidence scale.} \label{table:jeffreys}
\end{table}

A standard reference scale for the strength of evidence given by the
Bayes factor is the Jeffreys scale \cite{jeff}, shown in
Table~\ref{table:jeffreys}.

We compute the uncertainties in the Bayes
factor following a procedure described in Appendix~A.

The method presented above treats tracker behaviour as a Boolean
one-parameter property. It is thus insensitive to intrinsic biases
of the combined potential parameterization and parameter priors in
fulfilling the different tracker criteria, as well as how close to
the tracker criterion limits models typically fall. It would be
possible to go further and estimate the distributions of parameters
measuring each of the three tracker criteria. We outline a possible
procedure for this in Appendix~B, but present data do not appear to
justify such a sophisticated approach and we do not pursue this
further here.

\section{Results}

\subsection{Parameter estimation}

\begin{figure}[t]
\includegraphics[width=\linewidth]{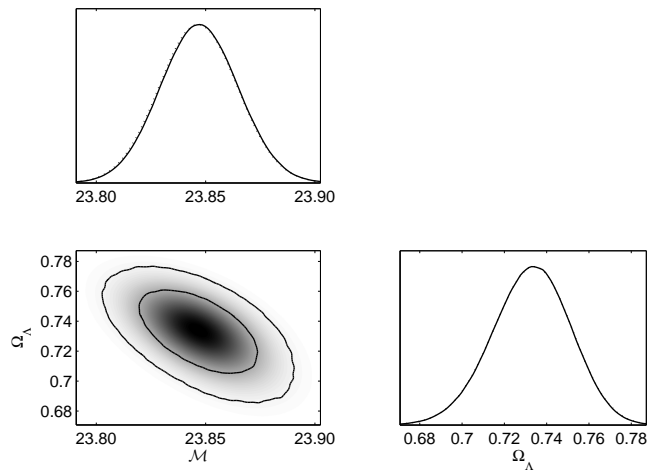}
\caption{One and two-dimensional likelihood distributions for a
cosmological constant model ($\Lambda$). Solid lines are
marginalized 1D likelihoods and dotted lines mean 1D likelihoods.
Solid 2D contours represent 68.3\% and 95.4\% regions of the
marginalized distribution, and shading reflects the mean
distribution.} \label{fig:lambda_tri}
\end{figure}

We present the probability distributions for the fitted models in
Figs.~\ref{fig:lambda_tri}--\ref{fig:m0n1_tri}. Marginalized
parameter constraints and best-fit values are given in Tables
\ref{tab:taylor} and \ref{tab:pade}. Plots of some dynamical
properties of the best-fit models can be found in
Figs.~\ref{fig:vphi} and \ref{fig:omegaphi}. The results are discussed
further below, and model comparison carried out in the following
subsection.

\subsubsection{Cosmological constant ($D=2$)}

The probability distributions for the cosmological constant case are
shown in Fig.~\ref{fig:lambda_tri}. The parameter constraints in Table
\ref{tab:taylor} are improved by roughly a factor of two compared to
our previous analysis \cite{oldppr}. They differ slightly from the
results of Ref.~\cite{liddledeev} using the same dataset, albeit
within uncertainties. This is most likely due to their different
treatment of SNLS SNIa errors.

\subsubsection{Skater ($D=3$)}

\begin{figure}[t]
\includegraphics[width=\linewidth]{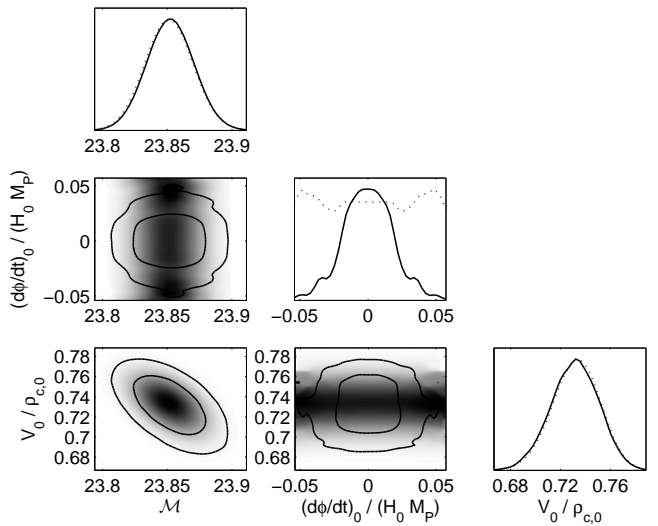}
\caption{As \reffig{fig:lambda_tri} for a `skater' model, a constant
potential with kinetic energy.} \label{fig:skater_tri}
\end{figure}

\begin{figure}[t]
\includegraphics[width=\linewidth]{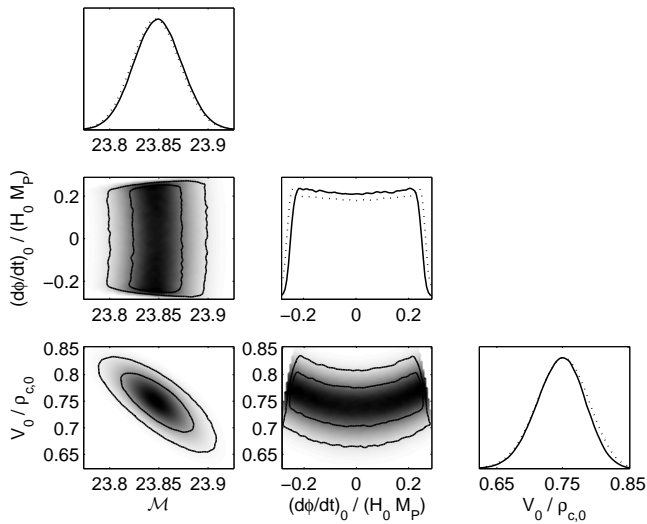}
\caption{As \reffig{fig:lambda_tri} for a constant potential with
kinetic energy. SNLS data only. Note that here the prior
$\Omega_{\rm kin}(z\ge1)<0.5$ is applied only up to $z=2$.}
\label{fig:snls_skater_tri}
\end{figure}

The likelihood distributions are shown in Fig.~\ref{fig:skater_tri}
for the full dataset, and in Fig.~\ref{fig:snls_skater_tri} for SNLS
alone.  Note the symmetry in $\dot{\phi}_0$, due to the dependence
only on $\dot{\phi}_0^2$. The degeneracy between $V_0$ and
$\dot{\phi}_0$ present in our previous analysis (where
$|\dot{\phi}_0|$ was positively correlated with $V_0$) is no longer
apparent with the full dataset, while still being visible if we use
supernovae alone. This degeneracy stems from the fact that with
supernovae we are really only sensitive to an effective quintessence
equation of state \cite{maordl, maorw}, which the data require to
be close to $-1$. Thus, increasing the kinetic energy of the field
must be compensated by an increase in potential energy to maintain
the same effective equation of state.

Additionally, the mild preference in the Riess \textit{et al.}  `gold'
data for a non-zero $\dot{\phi}_0$ is not present in the SNLS sample,
despite the $\dot{\phi}_0$--$V_0$ degeneracy being present.  Instead,
the likelihood distribution is essentially flat in
$\dot{\phi}_0$. This could be a reflection of the better
quality/homogeneity of the SNLS sample over Riess \textit{et al.}
(another possibility is the difference in redshift coverage).  In the
previous analysis, these two effects conspired to give a different
best-fit value of $V_0$ in the skater scenario ($V_0=0.74$) compared
to the cosmological constant (where $V_0 = \Omega_{\Lambda} =
0.69$). That we here do not feel the degeneracy is to some degree
linked to our prior limiting $\Omega_{\rm kin}(z\ge1)<0.5$ now being
applied to much higher redshifts, restricting the range of allowed
$\dot{\phi}_0$. However the new data do reduce the degeneracy
significantly on their own (we checked by doing the analysis without
the prior on $\Omega_{\rm kin}$).  Also, using only the SNLS data with
$\Omega_{\rm kin}(1\le z \le 2)<0.5$ (Fig.~\ref{fig:snls_skater_tri}),
the flatness of the distribution in $\dot{\phi}_0$ ensures that the
best-fit value of $V_0$ in that case is only marginally different from
that for the full analysis, even though the degeneracy is stronger.
These observations illustrate the need for good-quality data sensitive
to perturbation growth history (e.g.~weak lensing) to break the
$\dot{\phi}_0$--$V_0$ degeneracy.

\subsubsection{Linear potential ($D=4$)}

\begin{figure}[t]
\includegraphics[width=\linewidth]{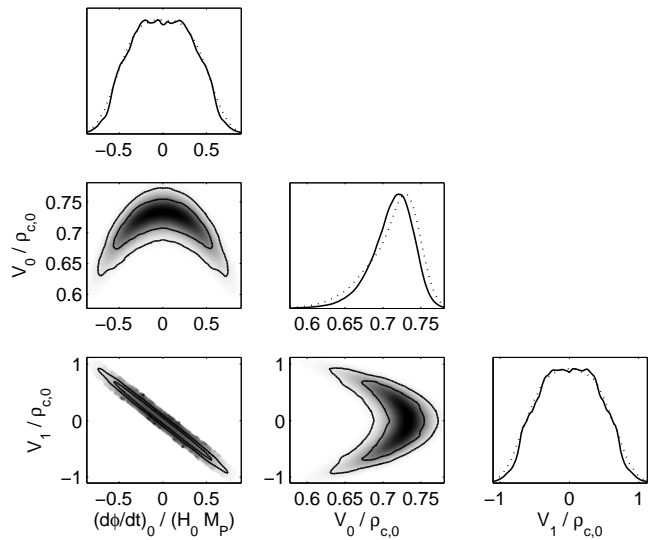}
\caption{As \reffig{fig:lambda_tri} for a linear potential.}
\label{fig:linear_tri}
\end{figure}

The likelihood distributions are shown in Fig.~\ref{fig:linear_tri}.
Note the bimodality of the $\dot{\phi}_0$--$V_1$ distribution,
reflecting that models are identical under simultaneous change of sign
of $\dot{\phi}_0$ and odd-order expansion coefficients. The first
change from previous constraints \cite{oldppr} is that the
$V_0$--$\dot{\phi}_0$ degeneracy is now clearly visible in the case of
the linear potential (there were only hints of it in the previous
analysis). That is to say, the data quality is getting closer to
hitting the degeneracy. In addition, we have a degeneracy between
$V_1$ and $\dot{\phi}_0$, coming from the possibility to achieve a
particular velocity of the field in the past by either changing the
present velocity $\dot{\phi}_0$ or the slope $V_1$.

Although not excluding the possibility, the new data do not favour a
potential where the field is rolling uphill (corresponding to the
upper right-hand and lower left-hand quadrants of the $\dot{\phi}_0 -
V_1$ distribution). This appears to be due to the new SNLS data, which
do not show a particular preference for a non-zero present field
velocity, thus not pushing us into these quadrants. It would appear
that the preference for an uphill rolling field found in our previous
analysis \cite{oldppr} was an artifact of the Riess \textit{et al.}
data. The observational consequences of such an uphill rolling field
could be interpreted as $w <-1$ if an `unsuitable'
parameterization is used to fit the data \cite{maorw, csakiwneg}. It
could thus be that the strong $w<-1$ preference found in the Riess
\textit{et al.}~data (see e.g.~Ref.~\cite{nessperi}) is due to some
systematic effect in the data, causing a preference for an uphill
rolling field and also corresponding to a preference for $w<-1$ in
fits of $w(z)$. This agrees with the findings of Nesseris and
Perivolaropoulos \cite{nessperi}, who for three different
parameterizations of $w$ find that the best-fit $w(z)$
consistently does not cross the phantom divide line $w = -1$
with the SNLS dataset, but does with the Riess \textit{et al.} `gold'
set. The analyses by Barger \textit{et al.} \cite{barger}, Xia
\textit{et al.} \cite{xia} and Jassal \textit{et al.} \cite{jassal}
lend support to this conclusion as well, as does a recent analysis by
Nesseris and Perivolaropoulos \cite{nessperi2}, who however find that
other cosmological data do gently favour phantom divide line crossing
provided $0.2 \lesssim \Omega_{\rm m} \lesssim 0.25$.

This also highlights the importance of interpreting analyses with
care, as we are not probing $w(z)$ directly \cite{maordl, maorw}.
This has been elaborated upon by several authors in terms of
eigenmodes, either as principal components \cite{pca} or weight
functions \cite{wtfun}.

\subsubsection{Pad\'e $R_{0/1}$ potential ($D=4$)}

\begin{figure}[t]
\includegraphics[width=\linewidth]{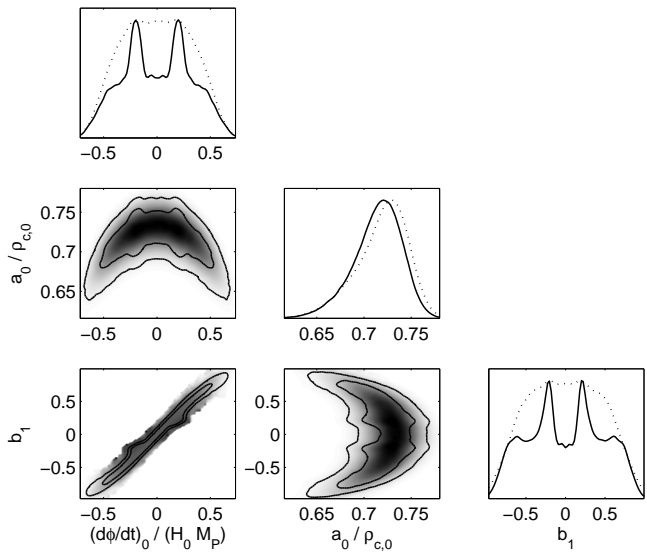}
\caption{As \reffig{fig:lambda_tri} for a Pad\'e series with
$M=0,N=1$ ($R_{0/1}$).} \label{fig:m0n1_tri}
\end{figure}

The likelihood distributions are shown in Fig.~\ref{fig:m0n1_tri}.  As
the $R_{0/1}$ potential is close to the linear case for small $\phi$,
we can use this to compare results.  That is, when $\dot{\phi}_0$ or
$b_1$ (which mainly determine the field velocity) are close to zero we
should expect results to compare well with the linear potential which,
comparing Fig.~\ref{fig:m0n1_tri} with Fig.~\ref{fig:linear_tri}, we
see they do.  Thus, the discussion above for the linear potential
applies to this case as well.  However, as we move away from
$\dot{\phi}_0=0$ and $b_1=0$, we see that $b_1$ is limited to somewhat
smaller values than for the linear case (using the relation $V_1
\approx -a_0 b_1$), while the constraints on $\dot{\phi}_0$ are almost
identical. This indicates that data prefer not to move very far away
from a linear potential. The other main feature of the likelihood
distributions are bumps found in the $\dot{\phi}_0$--$b_1$
distributions. These are a feature of the likelihood distribution, but
the exact size depends on our prior enforcing $\Omega_{\rm kin}(z\ge
1)<0.5$ up to high redshifts.

Pad\'e series, by construction, have poles. One might be concerned
about how this affects our results if the field reaches a pole, but
the data is sufficiently constraining that the poles are effectively
never felt. We tested this by doing the analysis with a prior
excluding all models where a pole is reached before $z=5$, and saw
no change in the results.

\begin{table*}
\begin{threeparttable}
  \begin{tabular}{|l|c|c|c|c|}
    \hline
   & \parbox{2.5cm}{\textbf{Cosmological} \\ \textbf{constant}
    $\mathbf{(\Lambda)}$} & \textbf{Skater} &
   \textbf{Linear} & \textbf{Quadratic} \tnote{a} ~~
\\
    \hline\hline
    $\mathcal{M}$ & $23.85^{+0.02}_{-0.02}$ & $23.86^{+0.01}_{-0.03}$ &
$23.86^{+0.02}_{-0.02}$ & $23.86$ \\
    \hline
     $\dot{\phi}_0 / H_0 M_{\rm P}$ & $-$ &
\parbox{5.2cm}{~\\$5.4\times10^{-5}$ ($5.5\times 10^{-2}$)\\
$|\dot{\phi}_0| / H_0 M_{\rm P} < 3.7\times10^{-2}$ ($95\%\, {\rm
CL}$)\\~} &
\parbox{4.5cm}{~\\$-2.7\times10^{-3}$ ($-6.5\times 10^{-2}$)\\
$|\dot{\phi}_0| / H_0 M_{\rm P} < 0.61$ ($95\%\, {\rm CL}$)\\~} &
$-0.15$\\
    \hline
    $V_0 / \rho_{\rm c,0}$ & $0.73^{+0.02}_{-0.02}$ &
$0.72^{+0.03}_{-0.01}$
&
$0.72^{+0.02}_{-0.03}$ & $0.73$ \\
    \hline
    $V_1 / \rho_{\rm c,0}$ & $-$ & $-$ &
    \parbox{4.5cm}{~\\$3.6\times10^{-3}$ ($8.7\times 10^{-3}$)\\
$|V_1| / \rho_{\rm c,0} < 0.76$ ($95\%\, {\rm
CL}$)\\~} & $0.58$ \\
    \hline
    $V_2 / \rho_{\rm c,0}$ & $-$ & $-$ & $-$ & $2.1$ \\
    \hline
    $- 2 \ln {\cal L}_{\rm max}$ & $113.6$ & $113.4$ & $113.4$ &
$112.9$ \\
    \hline
    $\mathrm{BIC}$ & $123.1$ & $127.7$ & $132.4$ & $136.7$ \\
      \hline
    $\mathrm{BIC} - \mathrm{BIC}_{\Lambda}$ & $0$ & $4.6$ & $9.3$ &
$13.6$\\
    \hline
  \end{tabular}
  \begin{tablenotes}
  \item[a] Since at least one parameter is unconstrained by the data
  for this model, we only give the best-fit parameter values found
  in our Markov chains.
  \end{tablenotes}
\end{threeparttable}
\caption{Marginalized median and best-fit model parameters and BIC
values for the cosmological constant ($\Omega_{\Lambda} =
V_0/\rho_{\rm c,0}$) and Taylor-series parameterizations. Best-fit
values are given in parentheses when differing from the median. Note
that the likelihood distribution is symmetric under simultaneous
change of sign of $\dot{\phi}_0$ and odd-order potential expansion
coefficients.} \label{tab:taylor}
\end{table*}

\begin{table*}
\begin{threeparttable}
  \begin{tabular}{|l|c|c|c|}
    \hline
    & \textbf{Pad\'e} $\mathbf{R_{0/1}}$ & \textbf{Pad\'e}
    $\mathbf{R_{0/2}}$ \tnote{a} ~~& \textbf{Pad\'e}
    $\mathbf{R_{1/1}}$ \tnote{a} ~~
\\
    \hline\hline
    $\mathcal{M}$ & $23.86^{+0.02}_{-0.02}$ & $23.86$ & $23.86$ \\
    \hline
     $\dot{\phi}_0 / H_0 M_{\rm P}$ &
\parbox{4.5cm}{~\\$1.2\times10^{-3}$ ($0.20$)\\ $|\dot{\phi}_0| / H_0
  M_{\rm P} < 0.57$ ($95\%\, {\rm
CL}$)\\~} & $-3.9\times 10^{-2}$ & $-9.8\times
  10^{-2}$ \\
    \hline
    $a_0 / \rho_{\rm c,0}$ & $0.72^{+0.02}_{-0.03}$ & $0.73$ & $0.73$ \\
    \hline
    $a_1 / \rho_{\rm c,0}$ & $-$ & $-$ & $-0.18$\\
    \hline
    $b_1$ & \parbox{3.5cm}{~\\$2.1\times10^{-3}$ ($0.18$)\\
$|b_1| < 0.82$ ($95\%\, {\rm
CL}$)\\~} & $-0.41$ & $-0.29$ \\
    \hline
    $b_2$ & $-$ & $-1.2$ & $-$\\
    \hline
    $- 2 \ln {\cal L}_{\rm max}$ & $113.3$ & $112.9$ & $113.3$ \\
    \hline
   $\mathrm{BIC}$ & $132.3$ & $136.7$ & $137.1$ \\
      \hline
    $\mathrm{BIC} - \mathrm{BIC}_{\Lambda}$ & $9.2$ & $13.6$ & $14.0$ \\
    \hline
  \end{tabular}
  \begin{tablenotes}
  \item[a] See Note a of Table~\ref{tab:taylor}.
  \end{tablenotes}
\end{threeparttable}
 \caption{Marginalized median and best-fit model
parameters and BIC values for the Pad\'e series parameterizations.
Best-fit values are given in parentheses when differing from the
median.} \label{tab:pade}
\end{table*}

\begin{figure}[t]
\includegraphics[width=0.95\linewidth]{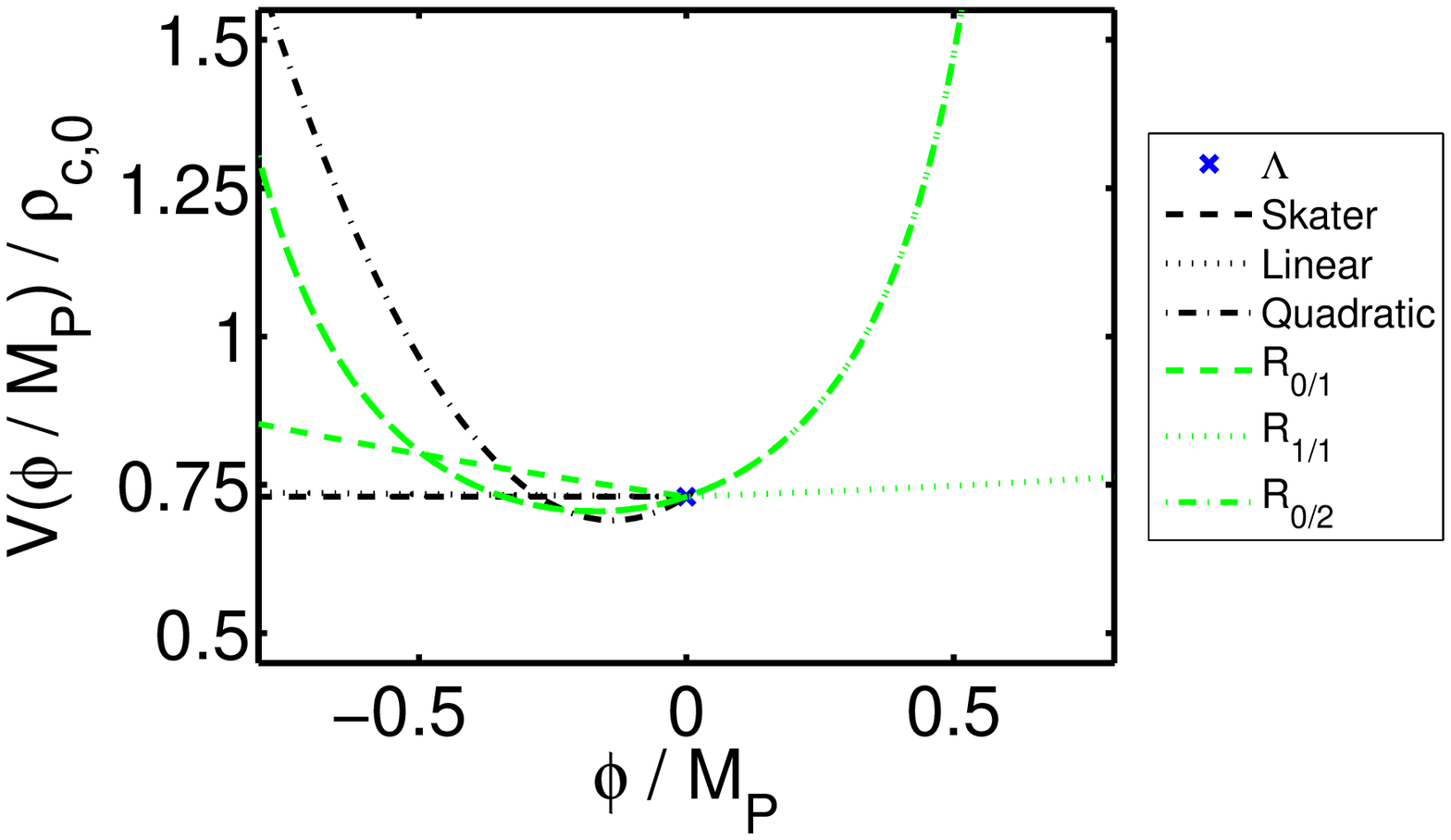}\\
\includegraphics[width=0.95\linewidth]{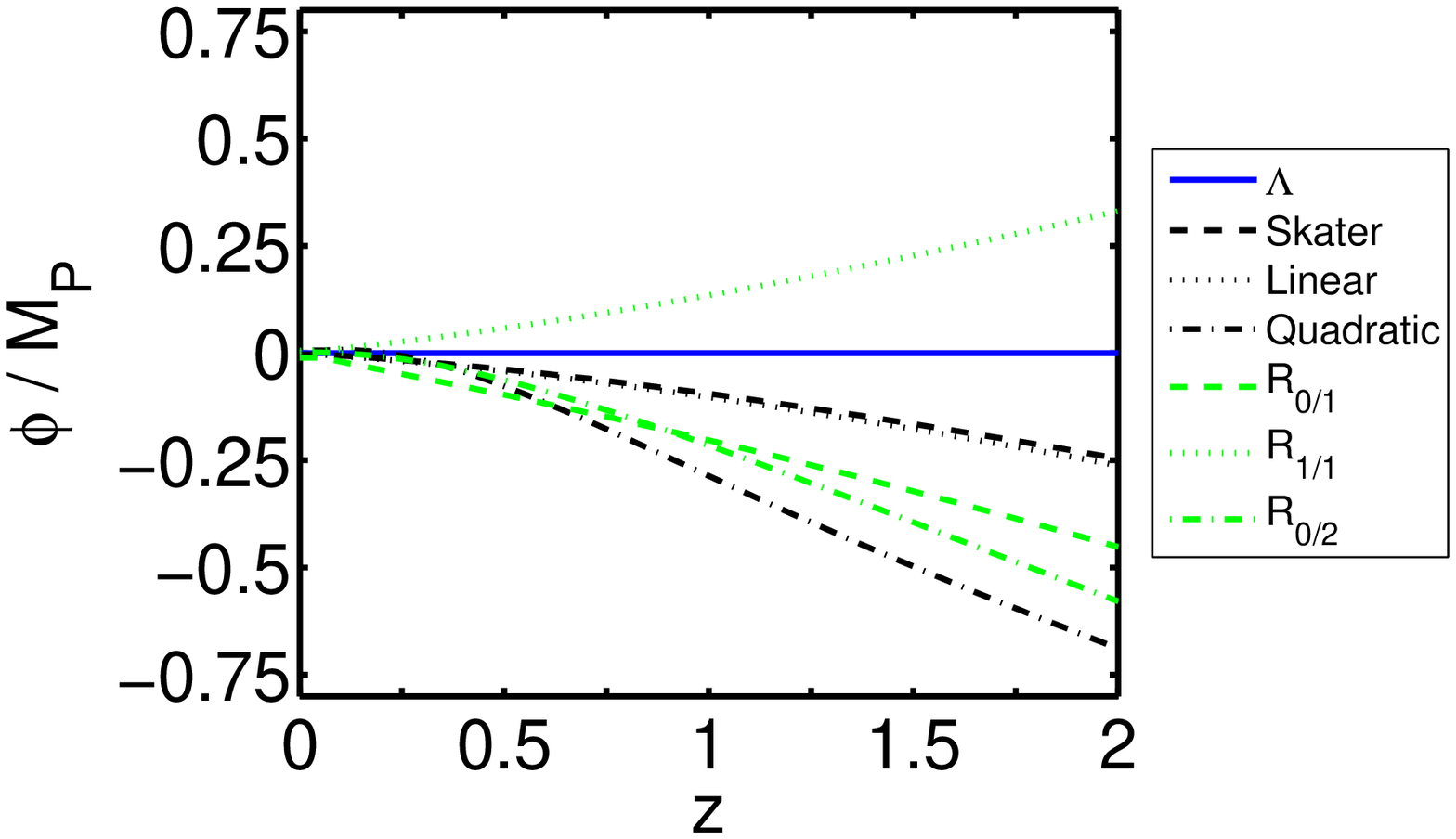}\\
\includegraphics[width=0.95\linewidth]{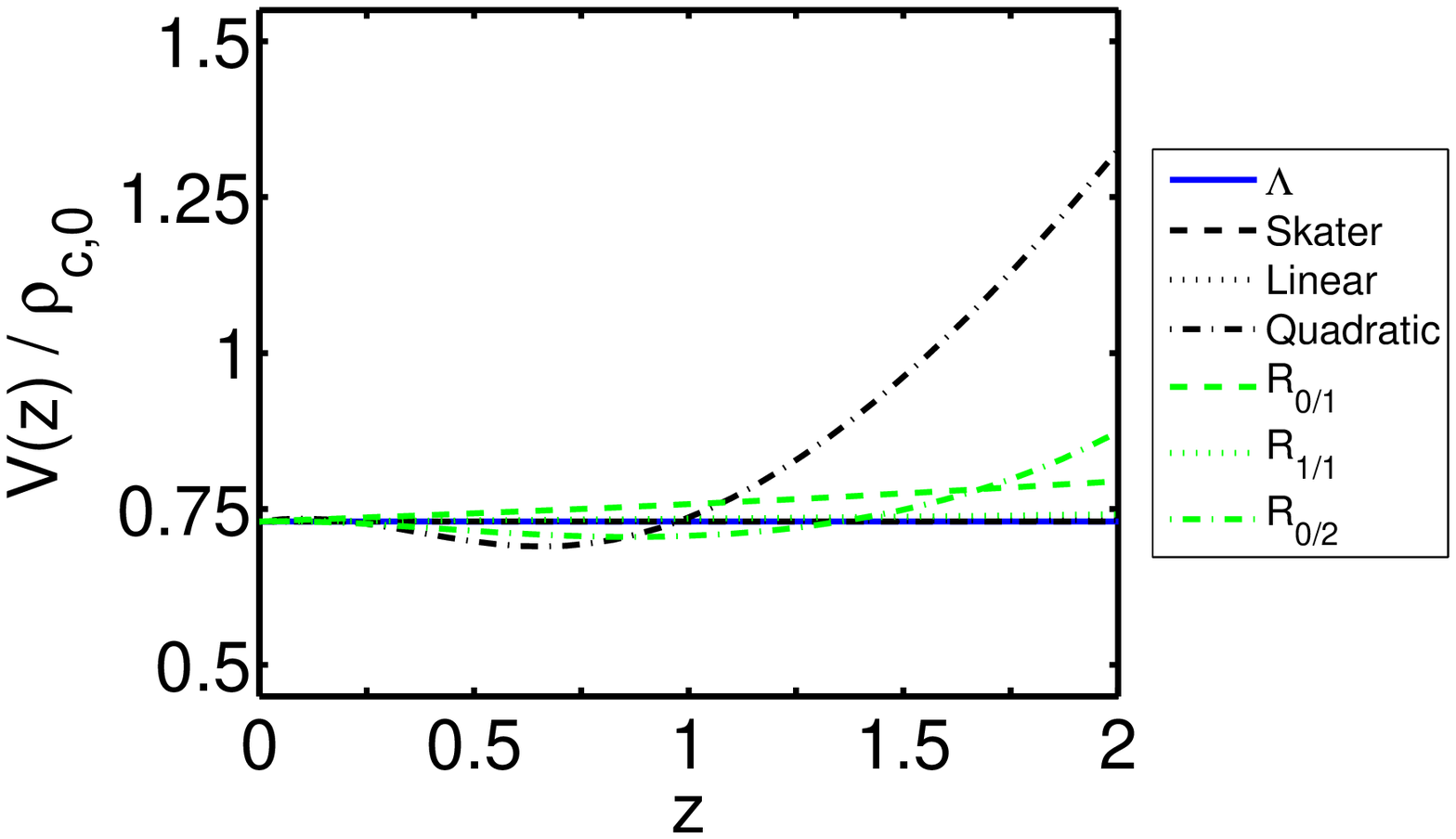}
\caption{Best-fit potentials as a function of the field, and the field
  and potentials as a function of redshift.}
\label{fig:vphi}
\label{fig:vz}
\end{figure}

\begin{figure}[t]
\includegraphics[width=0.95\linewidth]{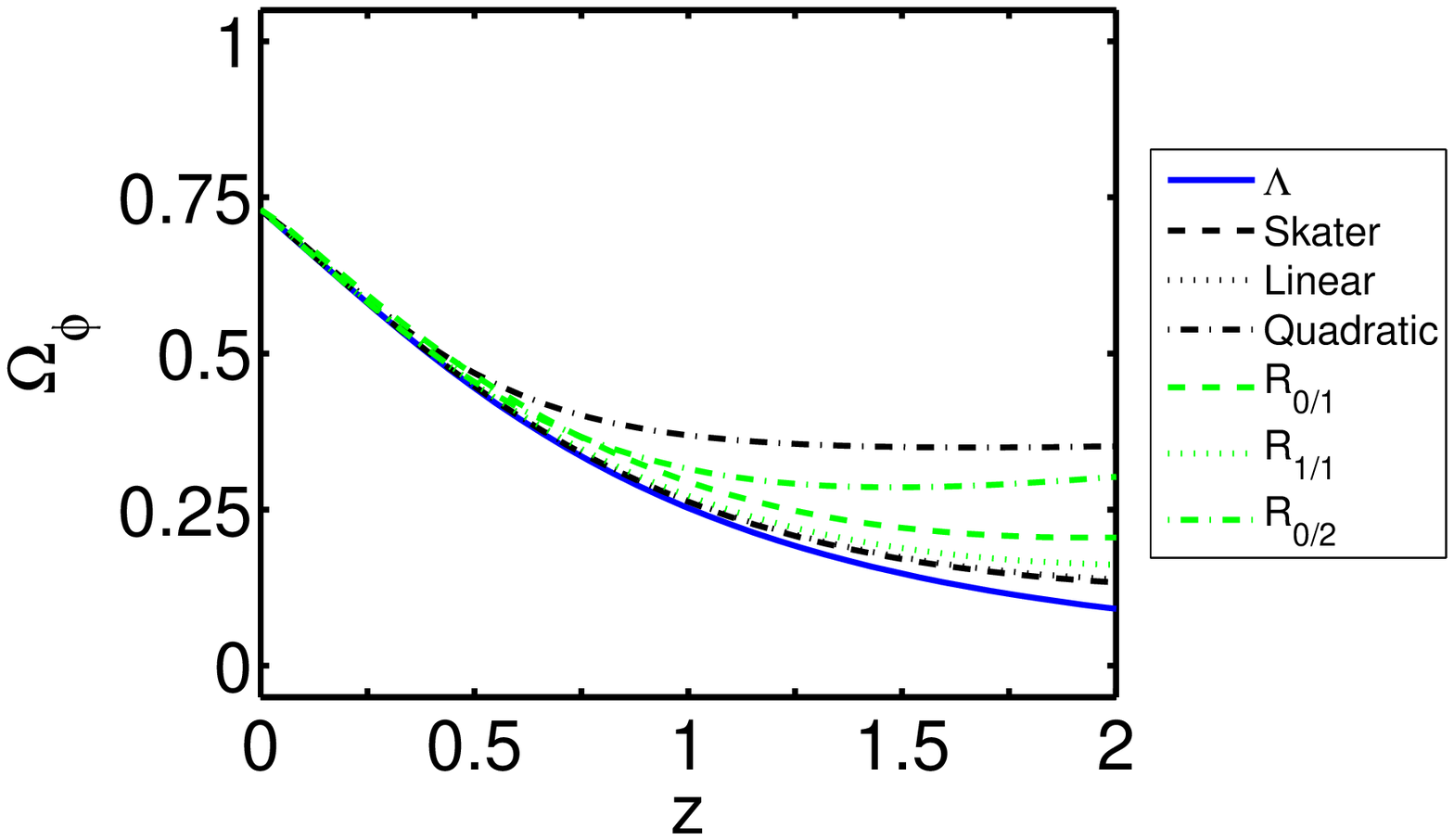}\\
\includegraphics[width=0.95\linewidth]{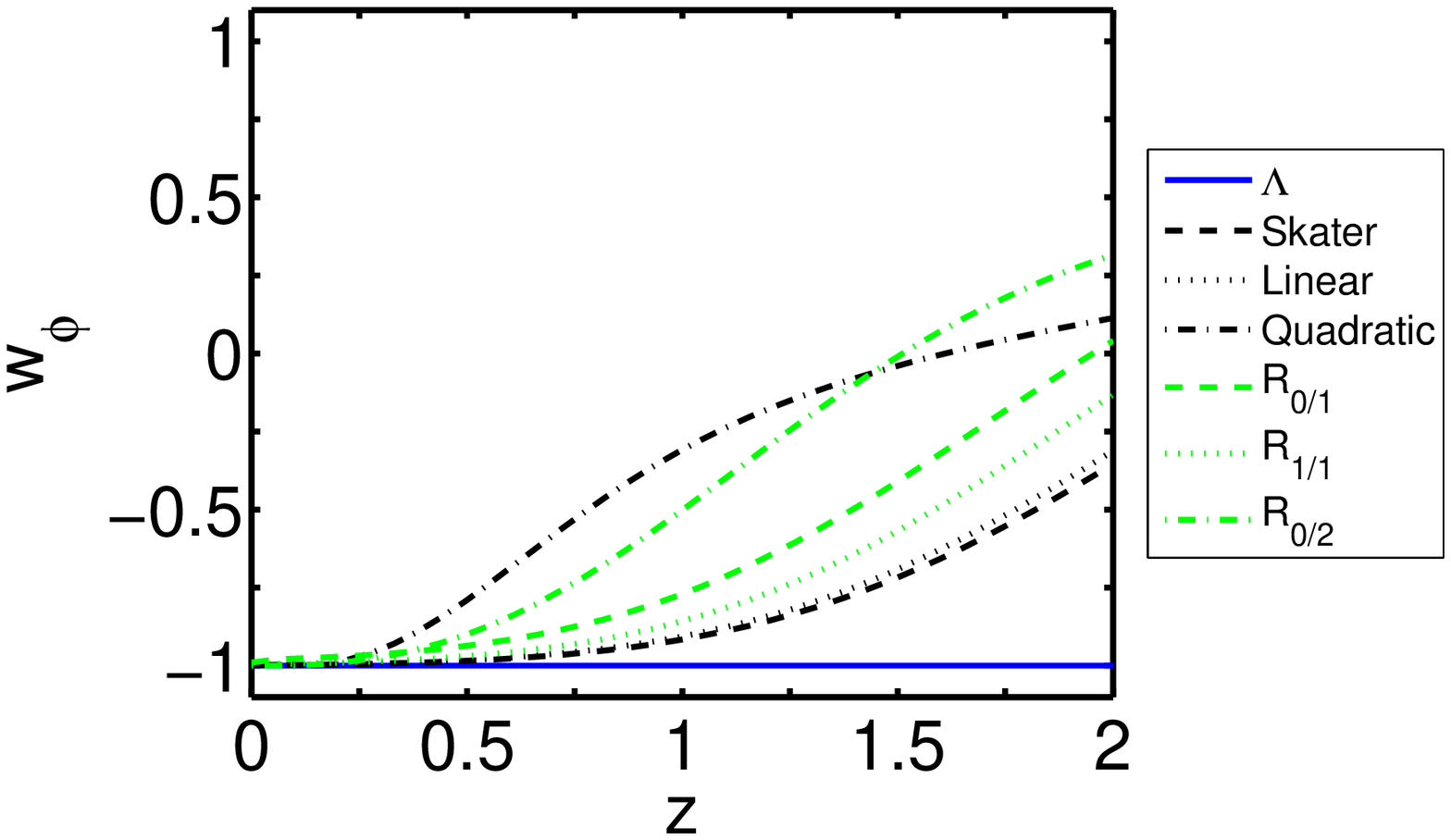}
\caption{Best-fit $\Omega_\phi$ and $w_\phi$ as a function of
  redshift.}
\label{fig:wphi}
\label{fig:omegaphi}
\end{figure}

\subsubsection{Models with $D>4$}

In the three cases with $D=5$ (quadratic, $R_{1/1}$, and $R_{0/2}$),
we find that the additional parameter is unconstrained by the data
and, as in Ref.~\cite{oldppr}, we learn nothing useful about
parameters from these models. Their principal interest lies in model
comparison, discussed next, where the best-fit found can still be used
to assess how the models compare in explaining the data.

\subsection{Model comparison}

The BIC values obtained for all models are shown in
Tables~\ref{tab:taylor} and \ref{tab:pade}. Note that although some
parameterizations have unconstrained parameters, their BIC value can
be evaluated with Eq.~(\ref{eq:bic}) from the best fit found in our
Monte Carlo Markov chains. It is clear that the cosmological
constant, showing a BIC difference of at least $4.6$ compared to the
other models, is positively favoured. This is a strengthening
compared to our previous analysis where this value was $4.0$. In
fact, the best-fit $\chi^2$ changes only marginally between models,
thus providing strong evidence against linear/Pad\'e $R_{0/1}$ and
higher-order potentials whose extra parameters add no value. An
interesting feature of the new dataset is that it much more strongly
disfavours a quadratic potential over the other Taylor expansions
than just the Riess \emph{et al.} data. Likewise, the lowest-order
Pad\'e expansion is favoured by the same amount compared to the
higher-order Pad\'e expansions.

The best-fit cosmologies (Figs.~\ref{fig:vphi} and \ref{fig:omegaphi})
now show more convergence in their dynamical properties, although
still exhibiting increasing variation with redshift. In particular, we
find that where previously the evolution of $\Omega_{\phi}$ for the
best-fit quadratic potential was such that $\Omega_{\phi}$ stayed
between $0.75$ and $0.96$ (for $0\le z\le 2$), the evolution is now
very reasonable (see Fig.~\ref{fig:omegaphi}). The strong evolution
previously seen in $w_{\phi}$ is now more limited, reflecting the
order-of-magnitude smaller best-fit values for $\dot{\phi}_0$ and
$V_1$ (though the overall compression of the uncertainties is much
less than this).

All best-fit models fall into the `freezing' category of Caldwell and
Linder \cite{CL}. For the skater model this behaviour is built-in, but
it is somewhat intriguing in terms of naturalness that the best-fit
linear potential exhibits freezing while at the same time rolling
downhill (see Figs.~\ref{fig:vz} and \ref{fig:wphi}).  The potentials
with curvature incorporate this best-fit behaviour by making the field
reach the potential minimum in the recent past (around $z=0.5$ to
$z=1$), thus providing a braking force to precipitate the accelerated
expansion of the universe. This situation would appear somewhat more
natural from a dynamical point of view, and it could be that the
best-fit linear potential is trying to approximate this, though data
is unable to sufficiently constrain the models with curvature in the
potential. On the other hand, model selection using the BIC also
strongly disfavours these models. The conclusion must be that
complementary or better-quality data is needed to resolve this
possible contradiction.

If the linear-potential results stand up, they will put the
well-motivated models of quintessence based on pseudo-Nambu-Goldstone
bosons (pNGBs) \cite{pngb} and similar models under pressure, as these
rely on a thawing field that is becoming dynamical and cosmologically
dominant in the present epoch. However a field just passing the
potential minimum fits well with the pNGB picture, as well as other
tracker-type potentials that show a cross-over behaviour, such as the
SUGRA \cite{sugra} and Albrecht--Skordis \cite{AS} potentials where
the field is starting to feel a curvature in the potential at late
times. These models exhibit early quintessence \cite{earlyquint}, and
can thus be constrained using big bang nucleosynthesis and CMB
observations \cite{bean}. It will be interesting to see what future
data, including those sensitive to perturbation growth and supernovae,
can tell.

These observations are in line with studies by e.g.~Bludman
\cite{bludman} and Linder \cite{linderpaths}, who both conclude that
quintessence generically cannot be described by slow-roll, and that
tracking must break down and move towards slow-roll in the recent past
(begging the question why this is happening precisely now).

\subsection{Tracker viability}

In carrying out the tracker viability analysis, we consider four
implementations in all by combining two choices of conditions. The
first is to demand either that the field remains in the tracker regime
until the present, or that it is allowed to break out of tracking
after a redshift of $z=1$. The second is to consider two different
upper limits for the redshift range where the field is required to be
in the tracker regime, namely $z=2$ and $z=10$; the former more or
less represents where the data actually lie, while the latter
extrapolates the potential to higher redshifts.

We find that all four cases give qualitatively the same outcome, and
so focus on just one choice, where tracking is imposed between $z=10$
and $z=1$.

\begin{figure}[t]
\includegraphics[width=0.95\linewidth]{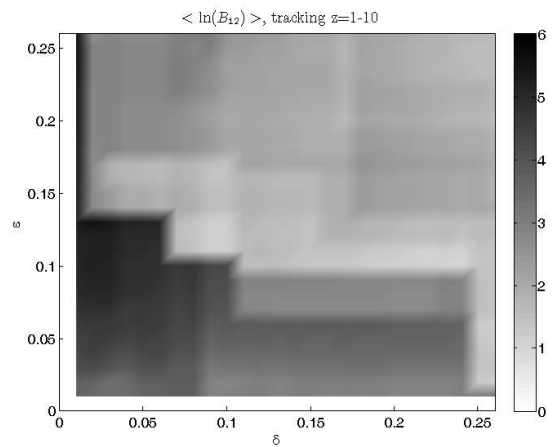}
\caption{Model average of $\ln(B_{12})$ for tracking required
between redshift 1 and 10, as a function of $\epsilon$ and
$\delta$.} \label{fig:track101av}
\end{figure}

The model average of $\ln B_{12}$, denoted $\langle \ln
B_{12}\rangle$, for this scenario is shown in
Fig.~\ref{fig:track101av}, for different combinations of $\epsilon$
and $\delta$. For combinations of sufficiently-small $\epsilon$ and
$\delta$, no models satisfying our tracker conditions are found in the
prior and/or posterior (with those $\epsilon$ and $\delta$ limits
different for the different parameterizations). We exclude these cases
from our model average, as they effectively correspond to an infinite
uncertainty in the derived value for $\ln B_{12}$. A very small
fraction of the models feel the presence of a pole at a redshift lower
than the upper tracker regime redshift, and are also excluded. We also
point out that for Pad\'e $R_{0/1}$, $\Gamma = 2$. Thus, the first two
tracker conditions are automatically fulfilled, corresponding to a
delta-function prior on $C_1$ and $C_2$ in the language of
Appendix~B. One might consider this a strong bias, and hence we
exclude this parameterization from our Bayes factor model average, and
thus use the quadratic, $R_{1/1}$ and $R_{0/2}$ potentials to arrive
at our conclusions.

It is clear from Fig.~\ref{fig:track101av} that the average indication
is in favour of tracker behaviour over non-tracker behaviour. The
smallest value of the Bayes factor in the figure is $0.98$. Limiting
our attention to the region where $\epsilon \le 0.1, \delta \le 0.1$,
and hence the tracker conditions are best obeyed, the smallest value
is $2.9$.  This general trend is seen in all four cases we analyze,
with the strongest preference for tracking in the case presented.
However, the model uncertainties in $\langle \ln B_{12} \rangle$ are
comparable to $\langle \ln B_{12} \rangle$ (particularly for small
$\epsilon$ and $\delta$) and a firm conclusion thus cannot be
drawn. (As a side note, the Poisson uncertainties are relatively small
and contribute at most on the order of $10\%$ to the total
uncertainties.)

The possible preference for tracker fields is in contrast with the
commonly-discussed expectation $w^{\rm eff}_{\phi} \gtrsim -0.8$ for
trackers, based on general inverse-power-law series potentials
\cite{steintrack} (here, $w^{\rm eff}_{\phi} = \int^1_{a_{\rm obs}}
w_{\phi}(a) \Omega_{\phi}(a) {\rm d}a / \int^1_{a_{\rm obs}}
\Omega_{\phi}(a) {\rm d}a$). While this seems to indicate that tracker
potentials are disfavoured by current data, our results suggest that
the data may act somewhat more strongly against non-tracker models
than against tracker ones.

\section{Conclusions}

We have updated parameter constraints on the quintessence potential
along with cosmological parameters using recent SNLS supernova
luminosity--redshift data, the WMAP3 CMB peak-shift parameter
measurement, and the SDSS measurement of baryon oscillations. The
preferred field dynamics appear robust under the different
parameterizations used.

We find that, compared to our previous work \cite{oldppr}, parameter
constraints are improved by roughly a factor of two. We also find that
linear-potential models where the field rolls uphill, although not
excluded, no longer provide the best fit to the data.  The previous
mild preference for these models appears to have been an artifact of
the Riess \textit{et al.}  `gold' SNIa data. This observation agrees
with the conclusions of other authors that the SNLS data do not
particularly favour an equation of state crossing the phantom divide
line, whereas the Riess \textit{et al.} data do.  Although
higher-order potentials are not constrained by the data, those
best-fit potentials exhibit `cross-over' behaviour, feeling a
curvature in the potential in the recent past. This qualitatively
agrees with some well-motivated tracking quintessence models.

{}From the point of view of model selection, the cosmological constant
is now even more strongly favoured compared to the dynamical models we
consider (see also Refs.~\cite{SWB,liddledeev}).  The models with
curvature in the potential are also strongly disfavoured as compared
to the constant and linear potentials, which appear dynamically less
natural in the context of the complete evolution expected from high
redshift.

We employ a model selection framework to investigate whether
potentials that exhibit tracker behaviour at intermediate/late times
are favoured by data over those potentials that do not. We conclude
that although our results show some indication that tracker behaviour
is favoured, the model uncertainty on the result is too large to draw
any firm conclusion.  We note that if the dynamics of our higher-order
best-fit potentials and the preference for a tracking potential both
stand up in the light of new data, the coincidence problem in the
context of quintessence may simply appear in a new guise --- why is
the field starting to slow-roll now?

It will be interesting to see how future perturbation growth data will
help break degeneracies, and, combined with supernova and CMB data,
constrain quintessence models and potentially change the model
selection picture as well.


\begin{acknowledgments}
M.S.\ was partially supported by the Stapley Trust and the Swedish
An\'er and Levin foundations, and in part by funds provided by
Syracuse University. A.R.L.\ and D.P.\ were supported by PPARC. M.S.\
thanks Mark Trodden and the Department of Physics, Syracuse
University, for generous hospitality during the completion of this
work. A.R.L.\ thanks the Institute for Astronomy, University of
Hawai`i, for hospitality while this paper was being completed. D.P.\
thanks the Department of Physics, University of Tokyo, for hospitality
while this paper was being completed. We thank Ariel Goobar and
Dominique Fouchez for helpful discussions, and Edvard M\"ortsell for
providing the SNLS data.  Partial analysis and plotting was made using
a modified version of {\tt GetDist} provided with {\tt CosmoMC}
\cite{Lewis:2002ah}. We additionally acknowledge use of the UK
National Cosmology Supercomputer funded by Silicon Graphics, Intel,
HEFCE and PPARC.
\end{acknowledgments}

\appendix

\section{Uncertainty in tracker Bayes factor estimates}

For simplicity of notation we define $E\equiv \ln B_{12}$ in this
Section. The uncertainty in our estimate of $E$ will consist of two
components: Poisson noise from sampling the distribution, and model
uncertainty. The Poisson noise goes as
\begin{eqnarray}
\sigma^2_{f_{\rm pri}} & = & f_{\rm pri}/N_{\rm pri} \,, \\
\sigma^2_{f_{\rm post}} & = & f_{\rm post}/N_{\rm post} \,,
\end{eqnarray}
where $N_{\rm pri}$ and $N_{\rm post}$ are the total numbers of
samples drawn from the prior and posterior distribution
respectively. Accordingly, using standard error propagation with
Eq.~(\ref{eq:bayes}), we have that
\begin{equation}
\sigma^2_{B_{12}} = D^2 \sigma^2_C + C^2 \sigma^2_D + 2B_{12}{\rm
cov}(C,D) \,,
\end{equation}
with $C = f_{\rm post}/(1-f_{\rm post})$ and $D = (1-f_{\rm
pri})/f_{\rm pri}$ so that $B_{12}=CD$. Additionally, we have
\begin{eqnarray}
 \sigma^2_C & = & \frac{\sigma^2_{f_{\rm post}}}{(1-f_{\rm post})^4}
 \,, \\
\sigma^2_D & = & \frac{\sigma^2_{f_{\rm pri}}}{f^4_{\rm pri}} \,.
\end{eqnarray}
In the absence of knowledge about the covariance between $C$ and
$D$, we can place an upper limit on the Poisson uncertainty,
\begin{equation}
\sigma^2_{B_{12}} \le \left(D \sigma_C + C \sigma_D \right)^2 \,.
\end{equation}
We use this upper limit as our estimate for the Poisson uncertainty.
The corresponding uncertainty in $E$ is then
\begin{eqnarray}
\nonumber
\sigma_{E} & = & \frac{\sigma_C}{C} + \frac{\sigma_D}{D} \\
 & = & \frac{\sigma_{f_{\rm post}}}{f_{\rm post}(1-f_{\rm post})} +
\frac{\sigma_{f_{\rm pri}}}{f_{\rm pri}(1-f_{\rm pri})} \,.
\end{eqnarray}

The model average of $E$ over $M$ models is given by (note that this
quantity is denoted $\langle \ln B_{12} \rangle$ in the main text)
\begin{equation}
\bar{E} = \frac{\sum_i E_i}{M}
\end{equation}
with an associated uncertainty
\begin{equation}
\sigma_{\bar{E}} = \sqrt{\frac{\sum_i (\bar{E} - E_i)^2}{M(M-1)}}
\,.
\end{equation}
We will now have an `error on the error' from the Poisson
uncertainty, given by
\begin{eqnarray}
\sigma_{\sigma_{\bar{E}}} = \sqrt{\frac{\sum_i (\bar{E} -
E_i)^2\sigma_{E_i}^2}{M(M-1)\sum_i (\bar{E} - E_i)^2}} \,,
\end{eqnarray}
so our final estimate of $E$ will be
\begin{eqnarray}
E & = & \bar{E} \\
 & \pm & \left[ \sqrt{\frac{\sum_i (\bar{E} - E_i)^2}{M(M-1)}} +
 \sqrt{\frac{\sum_i (\bar{E} - E_i)^2\sigma_{E_i}^2}{M(M-1)\sum_i
 (\bar{E} -  E_i)^2}} \right] \,. \nonumber
\end{eqnarray}

\section{Tracker probability distributions}

Here we briefly describe a possible extension of the tracker
analysis carried out in this paper, though we believe application to
present data would be premature.

To address the model uncertainty in the Bayes factor model average,
we consider the probability distributions of the parameters that
determine whether a model is classed as a tracker. In more detail,
we can define three different `tracker functions'
\begin{eqnarray}
C_1[\mathbf{z}_{\rm tr}] & = & \min_{\rm z \in \mathbf{z}_{\rm tr}}
\left( \Gamma(z) - 5/6 \right) \,, \\
C_2[\mathbf{z}_{\rm tr}] & = & \max_{\rm z \in \mathbf{z}_{\rm tr}}
\left|\Gamma(z)^{-1}\frac{d
  \Gamma(z)}{d\ln a}\right| \,, \\
C_3[\mathbf{z}_{\rm tr}] & = & \max_{\rm z \in \mathbf{z}_{\rm tr}}
\left|w_{\phi}(z)-w_{\rm tracker}(z)\right| \,,
\end{eqnarray}
where $\mathbf{z}_{\rm tr}$ is the redshift range for which the field
is required to exhibit tracker behaviour, and record their values for
all elements in our MCMC chains. Note that we do not include a
function corresponding to the constraint $w_{\phi}<0$, as $\max
w_{\phi}(z)$ will be a function of $C_1$ and $C_3$. From this we
obtain the posterior probability distribution $P(C_1, C_2, C_3 |
\Pi(\mathbf{\Theta}))$ given the prior distribution
$\Pi(\mathbf{\Theta})$ for our \emph{primary} cosmological parameters
$\mathbf{\Theta}$. Running the MCMC for the prior distribution as
well, we obtain the prior distribution $\Pi(C_1, C_2, C_3 |
\Pi(\mathbf{\Theta}))$.

We are then in a position to do \emph{importance sampling} (see
e.g.~Appendix B in Ref.~\cite{Lewis:2002ah} for a brief introduction)
using the prior and posterior we have calculated. We can change priors
for $C_1, C_2, C_3$ from those induced by $\Pi(\mathbf{\Theta})$ to
whichever we like and obtain the corresponding new posterior
distribution, since we only need to divide out the prior distribution
and multiply by the prior of our choice (with the exception of parts
of parameter space cut out by the primary prior $\Pi(\mathbf{\Theta})$
or very poorly sampled). A potential problem with this approach is
that optimal sampling of the posterior distribution in $C_1, C_2, C_3$
is not necessarily achieved by optimal sampling in the primary
parameters, and sufficient statistics may take a long time, i.e.  many
chain elements, to accumulate.

Setting natural priors for these new parameters may be
perceived as difficult (although not manifestly more arbitrary than
for other phenomenological parameterizations). A simple way of setting
the priors is to argue that we should be equally likely to draw a
parameter value that fulfils the corresponding tracker criterion, as
one that doesn't.  For instance, if we assume Gaussian priors, we get
\begin{eqnarray}
P(C_1) & = &
\frac{1}{\sqrt{2\pi}\sigma_{C_1}} \exp
\left[-\frac{C_1^2}{2\sigma^2_{C_1}} \right] \,, \\
P(C_2) & = & \frac{2}{\sqrt{2\pi}\sigma_{C_2}} \exp \left[
  -\frac{C_2^2}{2\sigma^2_{C_2}} \right] \theta(C_2)\,, \\
P(C_3) & = & \frac{2}{\sqrt{2\pi}\sigma_{C_3}} \exp \left[
-\frac{C_3^2}{2\sigma^2_{C_3}} \right] \theta(C_3) \,,
\end{eqnarray}
where $\theta$ is the Heaviside step function ($C_2$ and $C_3$ are
restricted to non-negative values by definition). The standard
deviations $\sigma_{C_2}$ and $\sigma_{C_3}$ are set by then
demanding
\begin{eqnarray}
\int_{C_2 \le \epsilon} P(C_2) {\rm d}C_2 & = & \int_{C_2 >
\epsilon} P(C_2)
{\rm d}C_2 \,, \\
\int_{C_3 \le \delta} P(C_3) {\rm d}C_3 & = & \int_{C_3 > \delta}
P(C_3) {\rm d}C_3 \,.
\end{eqnarray}
The case of $C_1$ is different, since we only have one inequality to
fulfil ($\Gamma>5/6$). Hence, we need to put a cut-off at some value
to determine the standard deviation. One could of course assign, for
example, flat priors in the same fashion.

Using this method, we can thus obtain a posterior distribution $P(C_1,
C_2, C_3)$ for a given prior distribution $\Pi(C_1, C_2, C_3)$ of our
choice, thus allowing a removal of correlation biases intrinsic to
particular parameterizations, which should reduce model
uncertainty. This method allows us to perform parameter estimation on
$C_1, C_2, C_3$ as well as model selection by calculating the Bayesian
evidence. It is of course applicable to general dynamical cosmological
properties one might wish to study. Carrying this out in practice can
however be involved since we might not be sampling efficiently in the
MCMC, and performing model selection in a robust manner would require
specialized code to address the sampling inefficiency and to handle
the use of a binned distribution.



\begin{thebibliography}{}

\bibitem{DETF} A. Albrecht \textit{et al.}~(Dark Energy Task Force report),
  \texttt{astro-ph/0609591}.

\bibitem{quint} C. Wetterich, Nucl. Phys. {\bf B302}, 668 (1988); B. Ratra
    and P. J. E. Peebles, Phys. Rev. D{\bf 37}, 3406
    (1988).

\bibitem{reviews} V. Sahni and A. A. Starobinsky, Int. J.
        Mod. Phys. \textbf{D9}, 373 (2000), \texttt{astro-ph/9904398};
        S. M. Carroll, Living Rev. Rel. {\bf 4}, 1 (2001);
        T. Padmanabhan, Phys. Rept.  {\bf 380}, 235 (2003), {\tt
        hep-th/0212290}; V. Sahni, Class. Quant. Grav. {\bf 19}, 3435
        (2002); T. Padmanabhan, Phys. Rept. {\bf 380}, 235 (2003);
        P. J. E. Peebles and B. Ratra, Rev. Mod. Phys. {\bf 75}, 559
        (2003); E. J. Copeland, M. Sami, and S. Tsujikawa (2006), {\tt
        hep-th/0603057}; V. Sahni and A. Starobinsky, {\tt
        astro-ph/0610026}.

\bibitem{rec} D. Huterer and M. S. Turner, Phys. Rev. D{\bf 60},
  081301(R) (1999), {\tt astro-ph/9808133};
    A. A. Starobinsky, JETP Lett.
     {\bf 68}, 757 (1998) [Pisma Zh. Eksp. Teor. Fiz. {\bf 68}, 721
     (1998)], {\tt astro-ph/9810431};
  T. Nakamura and T. Chiba,
  Mon. Not. Roy. Astron. Soc.  {\bf 306}, 696 (1999),
  \texttt{astro-ph/9810447};
    T. D. Saini, S. Raychaudhury, V. Sahni, and A. A. Starobinsky,
  Phys. Rev. Lett.  {\bf 85}, 1162 (2000),
  \texttt{astro-ph/9910231};
     B. F. Gerke and G. Efstathiou,
     Mon. Not. Roy. Astron. Soc. {\bf 335}, 33 (2002), {\tt
       astro-ph/0201336}; C. Wetterich, Phys. Lett. B{\bf 594}, 17
     (2004), {\tt astro-ph/0403289}; J. Simon, L. Verde, and
     R. Jimenez, Phys. Rev. D{\bf 71}, 123001 (2005), {\tt
       astro-ph/0412269};
     Z. K. Guo, N. Ohta, and Y. Z. Zhang,
  Phys. Rev. D{\bf 72}, 023504 (2005),
  \texttt{astro-ph/0505253};
      S. Tsujikawa,
  Phys. Rev. D{\bf 72}, 083512 (2005),
  \texttt{astro-ph/0508542};
  Z. K. Guo, N. Ohta, and Y. Z. Zhang,
  \texttt{astro-ph/0603109};
  X. Zhang, Phys. Rev. D{\bf 74}, 103505 (2006),
  \texttt{astro-ph/0609699}.

\bibitem{dalyrec}
  R. A. Daly and S. G. Djorgovski,
  Astrophys. J.  {\bf 597}, 9 (2003),
  \texttt{astro-ph/0305197};
 R. A. Daly and S. G. Djorgovski,
  Astrophys. J.  {\bf 612}, 652 (2004),
  \texttt{astro-ph/0403664};
  R. A. Daly and S. G. Djorgovski,
  \texttt{astro-ph/0512576};
  R. A. Daly and S. G. Djorgovski,
  \texttt{astro-ph/0609791}.

\bibitem{oldppr}
  M. Sahl\'en, A. R. Liddle, and D. Parkinson,
  Phys. Rev. D{\bf 72}, 083511 (2005),
  {\tt astro-ph/0506696}.

\bibitem{riess} A. G. Riess {\it et al.}  [Supernova Search Team
    Collaboration], Astrophys. J.  {\bf 607}, 665 (2004),
    {\tt astro-ph/0402512}.

\bibitem{wmap3}
  D. N. Spergel {\it et al.} [WMAP collaboration],
 \texttt{astro-ph/0603449}.

\bibitem{sdssbao}
  D. J. Eisenstein {\it et al.}  [SDSS Collaboration],
 Astrophys. J.  {\bf 633}, 560 (2005),
  \texttt{astro-ph/0501171}.

\bibitem{snls}
  P. Astier {\it et al.},
  Astron. Astrophys.  {\bf 447}, 31 (2006),
  \texttt{astro-ph/0510447}.

\bibitem{HP} D. Huterer and H. V. Peiris, \texttt{astro-ph/0610427}.

\bibitem{CL} R. R. Caldwell and E. V. Linder, Phys. Rev. Lett. {\bf
95}, 141301 (2005), \texttt{astro-ph/0505494}.

\bibitem{linderskater}
  E. V. Linder,
  Astropart. Phys.  {\bf 24}, 391 (2005),
  \texttt{astro-ph/0508333}.

\bibitem{pade} G. A. Baker and P. R. Graves-Morris, {\em Pad\'e
Approximants}, Vol. 1 \& 2, Addison-Wesley (1981).

\bibitem{linderhowmany}
  E. V. Linder and D. Huterer,
  Phys. Rev. D{\bf 72}, 043509 (2005),
  \texttt{astro-ph/0505330}.

\bibitem{Maor:2002rd} I. Maor and R. Brustein, Phys. Rev. D{\bf 67},
    103508 (2003), {\tt hep-ph/0209203}.

\bibitem{reftrack}
  P. G. Ferreira and M. Joyce,
  Phys. Rev. Lett.  {\bf 79}, 4740 (1997),
  \texttt{astro-ph/9707286};
  E. J. Copeland, A. R. Liddle, and D. Wands,
  Phys. Rev. D{\bf 57}, 4686 (1998),
  \texttt{gr-qc/9711068};
  P. G. Ferreira and M. Joyce,
  Phys. Rev. D{\bf 58}, 023503 (1998),
  \texttt{astro-ph/9711102};
  I. Zlatev, L. Wang, and P. J. Steinhardt,
  Phys. Rev. Lett.  {\bf 82}, 896 (1999),
  \texttt{astro-ph/9807002};
  A. R. Liddle and R. J. Scherrer,
  Phys. Rev. D{\bf 59}, 023509 (1998),
  \texttt{astro-ph/9809272};
  I. Zlatev and P. J. Steinhardt,
  Phys. Lett. B {\bf 459}, 570 (1999),
  \texttt{astro-ph/9906481};
  R. de Ritis, A. A. Marino, C. Rubano, and P. Scudellaro,
  Phys. Rev. D{\bf 62}, 043506 (2000),
  \texttt{hep-th/9907198};
  T. Barreiro, E. J. Copeland, and N. J. Nunes,
  Phys. Rev. D{\bf 61}, 127301 (2000),
  \texttt{astro-ph/9910214};
  P. Brax and J. Martin,
  \texttt{astro-ph/9912005};
  L. A. Urena-Lopez and T. Matos,
  Phys. Rev. D{\bf 62}, 081302 (2000),
  \texttt{astro-ph/0003364};
   V. B. Johri,
  Phys. Rev. D{\bf 63}, 103504 (2001),
  \texttt{astro-ph/0005608};
  R. Bean, S. H. Hansen, and A. Melchiorri,
  Phys. Rev. D{\bf 64}, 103508 (2001),
  \texttt{astro-ph/0104162};
  C. Rubano and J. D. Barrow,
  Phys. Rev. D{\bf 64}, 127301 (2001),
  \texttt{gr-qc/0105037};
  V. B. Johri,
  Class. Quant. Grav.  {\bf 19}, 5959 (2002)
  \texttt{astro-ph/0108247};
  C. Baccigalupi, A. Balbi, S. Matarrese, F. Perrotta, and N. Vittorio,
  Phys. Rev. D{\bf 65}, 063520 (2002),
  \texttt{astro-ph/0109097};
  W. Wang and B. Feng,
  Chin. J. Astron. Astrophys.  {\bf 3}, 105 (2003),
  \texttt{astro-ph/0508139};
   S. Tsujikawa,
  Phys. Rev. D{\bf 73}, 103504 (2006),
  \texttt{hep-th/0601178};
  L. Amendola, M. Quartin, S. Tsujikawa, and I. Waga,
  Phys. Rev. D{\bf 74}, 023525 (2006),
  \texttt{astro-ph/0605488};
  R. Das, T. W. Kephart, and R. J. Scherrer,
   \texttt{gr-qc/0609014}.

\bibitem{steintrack}
  P. J. Steinhardt, L. Wang, and I. Zlatev,
  Phys. Rev. D{\bf 59}, 123504 (1999),
  {\tt astro-ph/9812313}.

\bibitem{bludman}
  S. Bludman,
  Phys. Rev. D{\bf 69}, 122002 (2004),
  \texttt{astro-ph/0403526}.

\bibitem{linderpaths}
  E. V. Linder,
  Phys. Rev. D{\bf 73}, 063010 (2006),
  \texttt{astro-ph/0601052}.

\bibitem{wangkratochvil}
  Y. Wang, J. M. Kratochvil, A. Linde, and M. Shmakova,
  JCAP {\bf 0412}, 006 (2004),
  \texttt{astro-ph/0409264}.

\bibitem{fouchezpc}
  D. Fouchez, priv. comm.

\bibitem{peakshift}
  J. R. Bond, G. Efstathiou, and M. Tegmark,
 Mon. Not. Roy. Astron. Soc.  {\bf 291}, L33 (1997),
  \texttt{astro-ph/9702100};
  M. Zaldarriaga, D. N. Spergel, and U. Seljak,
  Astrophys. J.  {\bf 488}, 1 (1997),
  \texttt{astro-ph/9702157};
  G. Efstathiou and J. R. Bond,
  Mon. Not. Roy. Astron. Soc. {\bf 304}, 75 (1999),
  \texttt{astro-ph/9807103}.

\bibitem{wangr}
  Y. Wang and P. Mukherjee, Astrophys. J {\bf 650}, 1 (2006),
 \texttt{astro-ph/0604051}.

\bibitem{baoth}
  D. J. Eisenstein and W. Hu,
  Astrophys. J.  {\bf 496}, 605 (1998),
  \texttt{astro-ph/9709112};
  C. Blake and K. Glazebrook,
  Astrophys. J.  {\bf 594}, 665 (2003),
  \texttt{astro-ph/0301632}.

\bibitem{2dfbao}
  S. Cole {\it et al.}  [The 2dFGRS Collaboration],
 Mon. Not. Roy. Astron. Soc.  {\bf 362}, 505 (2005),
  \texttt{astro-ph/0501174}.

\bibitem{Schwarz} G. Schwarz, Annals of Statistics {\bf 5}, 461
  (1978).

\bibitem{Lid} A. R. Liddle, Mon. Not. Roy. Astron. Soc. {\bf 351}, L49
    (2004), {\tt astro-ph/0401198}.

\bibitem{jeff} H. Jeffreys, {\em Theory of Probability}, 3rd ed., Oxford
        University Press (1961).

\bibitem{Muk98} S. Mukherjee, E. D. Feigelson,  G. J. Babu, F. Murtagh, C.
    Fraley, and A. Raftery, Astrophys. J. {\bf 508}, 314 (1998),
    {\tt astro-ph/9802085}.

\bibitem{darkBIC} B. A. Bassett, P. S. Corasaniti, and M. Kunz,
  Astrophys. J. Lett. {\bf 617}, L1 (2004), {\tt astro-ph/0407364};
  M. Szydlowski and W. Godlowski, Phys. Lett. B{\bf 633}, 427 (2006);
  M. Szydlowski, A. Kurek, and A. Krawiec, Phys. Lett. B{\bf 642},
  171 (2006), {\tt astro-ph/0604327}.

\bibitem{bayesms}
  R. E. Kass and A. E. Raftery, Journal of the American
  Statistical Association {\bf 90}, 773 (1995);
  R. Trotta,
  \texttt{astro-ph/0504022};
  P. Mukherjee, D. Parkinson, P. S. Corasaniti, A. R. Liddle, and M. Kunz,
  Mon. Not. Roy. Astron. Soc.  {\bf 369}, 1725 (2006),
  \texttt{astro-ph/0512484};
  A. R. Liddle, P. Mukherjee, and D. Parkinson,
  \texttt{astro-ph/0608184}.

\bibitem{liddledeev}
  A. R. Liddle, P. Mukherjee, D. Parkinson, and Y. Wang,
  Phys. Rev. D{\bf 74}, 123506 (2006),
  \texttt{astro-ph/0610126}.

\bibitem{maorw}
  I. Maor, R. Brustein, J. McMahon, and P. J. Steinhardt,
  Phys. Rev. D{\bf 65}, 123003 (2002),
  \texttt{astro-ph/0112526}.

\bibitem{maordl}
  I. Maor, R. Brustein and P. J. Steinhardt,
  Phys. Rev. Lett.  {\bf 86}, 6 (2001)
  [Erratum-ibid.  {\bf 87}, 049901 (2001)],
  \texttt{astro-ph/0007297}.

\bibitem{csakiwneg}
  C. Csaki, N. Kaloper, and J. Terning,
  JCAP {\bf 0606}, 022 (2006),
  \texttt{astro-ph/0507148}.

\bibitem{nessperi}
 S. Nesseris and L. Perivolaropoulos,
  Phys. Rev. D{\bf 72}, 123519 (2005),
  \texttt{astro-ph/0511040}.

\bibitem{barger}
  V. Barger, E. Guarnaccia, and D. Marfatia,
  Phys. Lett. B {\bf 635}, 61 (2006),
  \texttt{hep-ph/0512320}.

\bibitem{xia}
  J. Q. Xia, G. B. Zhao, B. Feng, H. Li, and X. Zhang,
  Phys. Rev. D{\bf 73}, 063521 (2006),
  \texttt{astro-ph/0511625}.

\bibitem{jassal}
  H. K. Jassal, J. S. Bagla, and T. Padmanabhan,
  \texttt{astro-ph/0601389}.

\bibitem{nessperi2}
  S. Nesseris and L. Perivolaropoulos,
  \texttt{astro-ph/0610092}.

\bibitem{pca}
  D. Huterer and G. Starkman,
  Phys. Rev. Lett.  {\bf 90}, 031301 (2003),
  \texttt{astro-ph/0207517};
  D. Huterer and A. Cooray,
  Phys. Rev. D{\bf 71}, 023506 (2005),
  \texttt{astro-ph/0404062};
  R. G. Crittenden and L. Pogosian,
  \texttt{astro-ph/0510293};
  C. Stephan-Otto,
  Phys. Rev. D{\bf 74}, 023507 (2006),
  \texttt{astro-ph/0605403}.

\bibitem{wtfun}
  T. D. Saini, T. Padmanabhan, and S. Bridle,
  Mon. Not. Roy. Astron. Soc.  {\bf 343}, 533 (2003),
  \texttt{astro-ph/0301536};
  T. D. Saini,
  Mon. Not. Roy. Astron. Soc.  {\bf 344}, 129 (2003),
  \texttt{astro-ph/0302291};
  F. Simpson and S. Bridle,
  Phys. Rev. D{\bf 71}, 083501 (2005),
  \texttt{astro-ph/0411673};
  F. Simpson and S. Bridle,
  Phys. Rev. D{\bf 73}, 083001 (2006),
  \texttt{astro-ph/0602213}.

\bibitem{pngb}
  C. T. Hill, D. N. Schramm, and J. N. Fry,
  Comments Nucl. Part. Phys.  {\bf 19}, 25 (1989);
  J. A. Frieman, C. T. Hill, and R. Watkins,
  Phys. Rev. D{\bf 46}, 1226 (1992);
  M. Fukugita and T. Yanagida,
  YITP-K-1098 (1994);
  J. A. Frieman, C. T. Hill, A. Stebbins, and I. Waga,
  Phys. Rev. Lett.  {\bf 75}, 2077 (1995),
  \texttt{astro-ph/9505060};
  N. Kaloper and L. Sorbo,
  JCAP {\bf 0604}, 007 (2006),
  \texttt{astro-ph/0511543}.

\bibitem{sugra}
  P. Binetruy,
  Phys. Rev. D{\bf 60}, 063502 (1999),
  \texttt{hep-ph/9810553};
  P. Brax and J. Martin,
  Phys. Lett. B{\bf 468}, 40 (1999),
  \texttt{astro-ph/9905040};
  P. Brax, J. Martin and A. Riazuelo,
  Phys. Rev. D{\bf 64}, 083505 (2001),
  \texttt{hep-ph/0104240}.

\bibitem{AS}
  A. Albrecht and C. Skordis,
  Phys. Rev. Lett.  {\bf 84}, 2076 (2000),
  \texttt{astro-ph/9908085};
  C. Skordis and A. Albrecht,
  Phys. Rev. D{\bf 66}, 043523 (2002),
  \texttt{astro-ph/0012195}.

\bibitem{earlyquint}
  C. Wetterich,
  JCAP {\bf 0310}, 002 (2003),
  \texttt{hep-ph/0203266};
    C. Wetterich,
  Phys. Lett. B{\bf 561}, 10 (2003), \texttt{hep-ph/0301261};
  R. R. Caldwell, M. Doran, C. M. Mueller, G. Schaefer, and C. Wetterich,
  Astrophys.  J.  {\bf 591}, L75 (2003), \texttt{astro-ph/0302505};
  M. Doran and G. Robbers,
  JCAP {\bf 0606}, 026 (2006),
  \texttt{astro-ph/0601544}.

\bibitem{bean} R. Bean, S. H. Hansen, and A. Melchiorri, Phys. Rev.
  D{\bf 64}, 103508 (2001), \texttt{astro-ph/0104162}.



\bibitem{SWB} T. D. Saini, J. Weller, and S. L. Bridle, Mon. Not. Roy.
  Astron. Soc. {\bf 348}, 603 (2004), {\tt astro-ph/0305526}.

\bibitem{Lewis:2002ah} A. Lewis and S. Bridle, Phys. Rev. D{\bf 66},
    103511 (2002), {\tt astro-ph/0205436}.

\end{thebibliography}
\end{document}